\documentclass[prd,aps,floats,preprintnumbers,preprint,superscriptaddress,floatfix,nofootinbib]{revtex4}
\usepackage{graphicx, epsfig}
\newcommand{ \slashchar }[1]{\setbox0=\hbox{$#1$}   
   \dimen0=\wd0                                     
   \setbox1=\hbox{/} \dimen1=\wd1                   
   \ifdim\dimen0>\dimen1                            
      \rlap{\hbox to \dimen0{\hfil/\hfil}}          
      #1                                            
   \else                                            
      \rlap{\hbox to \dimen1{\hfil$#1$\hfil}}       
      /                                             
   \fi}                                             %

\def\ptmiss{\slashchar{p}_{T}}
\def\etmiss{\slashchar{E}_{T}}
\def\lsim{\mathrel{\raise.3ex\hbox{$<$\kern-.75em\lower1ex\hbox{$\sim$}}}}
\def\gsim{\mathrel{\raise.3ex\hbox{$>$\kern-.75em\lower1ex\hbox{$\sim$}}}}
\def\vp{{\vec p}}
\def\pt{p_T^{}}
\def\et{E_T^{}}
\def\cM{{\cal M}}
\def\cO{{\cal O}}
\def\ecm{E_{CM}}
\def\epem{e^+e^-}

\def\mev{{\rm MeV}}
\def\gev{{\rm GeV}}
\def\tev{{\rm TeV}}
\def\pb{{\rm pb}}
\def\fb{{\rm fb}}

\def\nbi{{\rm nb}^{-1}}
\def\fbi{{\rm fb}^{-1}}
\def\be{\begin{equation}}
\def\ee{\end{equation}}
\def\bea{\begin{eqnarray}}
\def\eea{\end{eqnarray}}

\begin{document}

\vskip 1cm
 \preprint{
 {\vbox{
 \hbox{MADPH--05--1434}
 \hbox{hep-ph/0508097}}}}

\title{COLLIDER PHENOMENOLOGY\\
Basic Knowledge and Techniques\footnote{I will maintain an updated 
version of these lectures at http://pheno.physics.wisc.edu/$\sim$than/collider-2005.pdf.}}

\author{ Tao Han }
\email{than@physics.wisc.edu}
\affiliation{Department of Physics, 1150 University Avenue,
University of Wisconsin, Madison, WI 53706, USA}


\begin{abstract}
This is the writeup for TASI--04 lectures on Collider Phenomenology.
These lectures are meant to provide an introductory  presentation 
on the basic knowledge and techniques for collider physics.
Special efforts have been made  for  those 
theorists who need to know some experimental  issues in 
collider environments, 
and for those experimenters who would like to know more about 
theoretical considerations in searching for new signals at colliders. 
\end{abstract}

\maketitle


\tableofcontents

\section{Introduction}

For the past several decades, high energy accelerators and colliders 
have been our primary tools for discovering new particles and for testing 
our theory of fundamental interactions. With the
expectation of the Large Hadron Collider (LHC) in mission in 2007,
and the escalated preparation for the International Linear Collider (ILC),
we will be fully exploring the physics at the electroweak scale and beyond
the standard model (SM) of the strong and electroweak interactions 
in the next twenty years. New exciting discoveries are highly anticipated 
that will shed light on the mechanism for electroweak symmetry breaking,
fermion mass generation and their mixings, on new fundamental symmetries 
such as Supersymmetry (SUSY) and grand unification of forces (GUTs),
even on probing the existence of extra spatial dimensions
or low-scale string effects, and on related cosmological implications
such as particle dark matter, baryon and CP asymmetries of the
Universe, and dark energy as well.

Collider phenomenology plays a pivotal role in building the 
bridge between theory and experiments. On the one hand,  one would like
to decode the theoretical models and to exhibit their experimentally 
observable consequences. On the other hand, one needs to interpret 
the data from experiments and to understand their profound implications. 
Phenomenologists working in this exciting era would naturally need to acquaint
both fields, the more the better. 

These lectures are aimed for particle physicists who need to know 
the basics in collider phenomenology, both experimental issues and
theoretical approaches.  Special efforts have been made for 
those theorists  who need to know some  realistic experimental issues
at high-energy collider environments,  and those experimenters 
who would like to know more about theoretical considerations 
in searching for generic new signals. In preparing these lectures, 
I had set up a humble goal. I would not advocate
a specific theoretical model currently popular or of my favorite; nor  would I
summarize the ``new physics reach"  in a model-dependent parameter space
at the LHC and ILC; nor would I get into a sophisticated level of experimental
simulations of detector effects.  
The goal of these lectures is to present the basic knowledge 
in collider physics including experimental concerns, 
and to discuss generic techniques for collider 
phenomenology  hopefully in a pedagogical manner. 

In Sec.~\ref{collider}, we first present basic collider parameters relevant to
our future phenomenological considerations. We then separately discuss 
$\epem$ linear colliders and hadron colliders for the calculational framework,
and for physics expectations within the SM. 
In Sec.~\ref{detector}, we discuss issues for particle detection  $-$
what do the elementary particles in the SM theory look like in a realistic
detector? $-$ which
are necessary knowledge but have been often overlooked by theory students. 
We also illustrate  what parameters of a detector and what measurements
should be important for a phenomenologist to pay attention to.  
Somewhat more theoretical topics are presented in Sec.~\ref{uncover}, 
where I emphasize a few important kinematical observables and suggest 
how to develop your own skills to uncover fundamental
dynamics from experimentally accessible kinematics. If I had more time
to lecture or to write,
this would be the section that I'd like to significantly expand. 
Some useful technical details are listed in a few Appendices. 

The readers are supposed to be familiar with the standard model of the
strong and electroweak interactions, 
for which I refer to Scott Willenbrock's lectures \cite{scott-sm}
and some standard texts \cite{books-sm}. 
I also casually touch upon topics in theories such as SUSY, 
extra dimensions, and new electroweak
symmetry breaking scenarios, for which I refer the readers to the 
lectures by Howie Haber \cite{howie-susy,PDG} on SUSY, and some recent 
texts \cite{books-susy}, 
Raman Sundrum \cite{raman-extrad} and Csaba Csaki \cite{csaba-higgs}
on physics with extra-dimensions. 
For more extensive experimental issues, 
I refer to Heidi Schellman's lectures \cite{exp}. 
The breath and depth covered in these lectures are obviously very limited.
For the readers who need more theoretical knowledge  on collider phenomenology,
there are excellent text books \cite{bp,esw} as references.
As for experimental issues, one may find a text \cite{book-exp} very
useful, or consult with the Technical Design Reports
(TDR) from various detector collaborations \cite{ATLAS,CMS,ILC}.

\section{High Energy Colliders: Our Powerful Tools}
\label{collider}

\subsection{ Collider Parameters}

In the collisions of two particles of masses $m_1$ and $m_2$
and momenta $\vp_1$ and $\vp_2$, the total energy squared in the
center-of-momentum frame (c.m.)  can be expressed in terms of a  
Lorentz-invariant Mandelstam variable (for more details, see
Appendix \ref{app-ps})
\bea
\nonumber
s  \equiv  (p_1+p_2)^2 = \left\{  
\begin{array}{ll}
(E_1 + E_2 )^2 \qquad   {\rm  in\  the\  c.m.\ frame}\ \vp_1+\vp_2=0, & \\
  m_1^2 + m_2^2 + 2(E_1E_2-\vp_1\cdot \vp_2). &  
\end{array}
\right.
\eea
In high energy collisions of our current interest, the beam particles are ultra-relativistic
and the momenta are typically much larger than their masses.
The total c.m.~energy of the two-particle system can thus be approximated as 
\bea
\nonumber
\ecm \equiv \sqrt s \hskip-0.2cm & \approx &\hskip-0.2cm  
\left\{  
\begin{array}{ll}
2E_1 \approx 2 E_2   &   {\rm  in\ the\ c.m.\ frame}\ \vp_1+\vp_2=0, \\
\sqrt{2 E_1 m_2} &   {\rm in\ the\ fixed\ target\  frame\ \vp_2=0}. 
\end{array}
\right.
\eea
while the kinetic energy of the system is
$T \approx  E_1$ in  the  fixed-target  frame $\vp_2=0$, 
and $T = 0$  in the  c.m.~frame $\vp_1+\vp_2=0.$
We see that only in the c.m.~frame, there will be no kinetic motion of
the system, and  the beam energies are maximumly 
converted to reach a higher threshold. This is the designing 
principle for colliders like LEP I,  LEP II and LHC at CERN; the SLC at 
SLAC; and the Tevatron at the Fermi National Accelerator Laboraroty. 
Their c.m.~energies are listed in Tables \ref{eecolliders} 
and \ref{hcolliders}, respectively. 

\begin{table}[tb]
    \begin{tabular}[t]{|c| c | c | c |c|c|c|}
       \hline
        Colliders  & $\sqrt s$ (GeV) & ${\cal{L}}$ & $\delta E/E$ & $f$& 
        polar. & L \\
 &(GeV) & (cm$^{-2}$s$^{-1}$)& & (kHz) & &(km) \\
        \hline        \hline
        LEP I   & $M_Z$ & $2.4\times10^{31}$  &  $\sim 0.1\%$   &  45   & $55\%$ & 26.7\\
        SLC    & $ \sim 100 $             & $2.5\times 10^{30}$  &  $0.12\%$ & 0.12 &  80\% & 2.9\\
        LEP II  & $\sim 210$ &  10$^{32}$   & $\sim 0.1\%$ & 45 & & 26.7 \\
\hline\hline
 &(TeV) & & & (MHz) & & \\
        ILC    & 0.5$-$1  & $2.5\times 10^{34}$ & $0.1\%$ &  3 & $80,60\%$ & 14-33\\
\hline
                CLIC    & 3$-$5  & $\sim 10^{35}$ & $0.35\%$ & 1500 & $80,60\%$ & 33-53\\
\hline
    \end{tabular}
\caption{Some $e^+e^-$ colliders and their important parameters:
c.m.~energy, instantaneous peak luminosity, relative beam energy spread, 
bunch crossing frequency, longitudinal beam 
polarization, and the total length of the collider. The parameters are mainly from
PDG {\protect \cite{PDG}}, 
ILC working group reports {\protect \cite{ILC}}, 
 and a recent CLIC report  {\protect \cite{CLIC}}.
}
\label{eecolliders} 
\end{table}

The limiting factor to the collider energy is the energy loss 
during the acceleration, known as the synchrotron radiation.
For a circular machine of radius $R$, the energy loss per revolution
is \cite{PDG,book-exp}
\bea
\label{synch}
\Delta E \propto {1\over R} \left({E\over m}\right)^4,
\eea
where $E$ is beam energy, $m$ the particle mass (thus $E/m$ is the
relativistic $\gamma$ factor). It becomes clear that an accelerator is more
efficient for a larger radius or a more massive particle.

In $\epem$ annihilations, the c.m.~energy may be fully converted into reaching the 
physics threshold. In hadronic collisions, only a fraction of the total c.m.~energy
is carried by the fundamental degrees of freedom, the quarks and gluons (called partons).
For instance, the Tevatron, with the highest c.m.~energy
available today, may reach an effective parton-level energy of a few hundred GeV;
while the LHC will enhance it to multi-TeV.

\begin{table}[tb]
    \begin{tabular}[t]{|c| c | c | c |c|c|c|}
       \hline
        Colliders  & $\sqrt s$& ${\cal{L}}$& $\delta E/E$ & $f$ & $\#$/bunch & L\\
          & (TeV) &  (cm$^{-2}$s$^{-1}$)&  & (MHz)& $(10^{10})$ & (km)\\
        \hline        \hline
        Tevatron & 1.96   &     $2.1\times 10^{32}$   &  $9\times 10^{-5}$  & 2.5 & $p$: 27, 
        $\bar p$: 7.5& 6.28\\
        HERA &  314  & $1.4\times 10^{31}$   &  $0.1,0.02\%$ &10  & $e$: 3, $p$: 7& 6.34 \\
        \hline    \hline
        LHC & 14   &   $10^{34}$   & $0.01\%$ & 40 & 10.5& 26.66\\        
        \hline
         SSC & 40   &   $10^{33}$   & $5.5\times 10^{-5}$ & 60 & 0.8 & 87\\        
        VLHC    & 40$-$170  & $2\times10^{34}$ & $4.4\times 10^{-4}$ & 53 & 2.6 & 233 \\
\hline
    \end{tabular}
\caption{Some hadron colliders and their important parameters 
{\protect \cite{PDG}}: c.m.~energy, instantaneous
peak luminosity, relative beam energy spread, bunch crossing frequency, 
number of particles per bunch, and the total length of the collider. For
reference, the cancelled SSC and a recently discussed future VLHC {\protect \cite{vlhc}}
are also listed. }
\label{hcolliders} 
\end{table}

\begin{figure}[h]
\psfig{figure=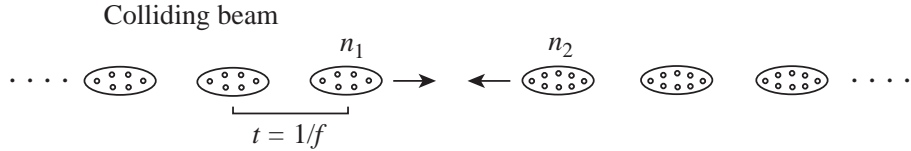,height=0.84in}
\caption{Colliding beams  with a bunch crossing frequency $f$.
\label{fig:collider}}
\end{figure}

Another important parameter for a collider is the instantaneous luminosity, 
the number of particles passing each other  per unit time through unit 
transverse area at the interaction point.  In reality, 
the particle  beams usually come in bunches, as roughly illustrated in
Figure \ref{fig:collider}. If there are $n_1$ particles in each bunch
in beam 1 and $n_2$ in each bunch in beam 2, then the collider luminosity scales as
\be
{\cal{L}} \propto f n_1 n_2/a,
\ee
where $f$ is beam crossing frequency and $a$ the transverse profile of the 
beams. The instantaneous luminosity is usually given in units of cm$^{-2}$ s$^{-1}$. 

The reaction rate, that is the number of scattering events per unit time,
is  directly proportional to the luminosity and is  given by\footnote{There 
will be another factor $\epsilon<1$ on the right-hand side, 
which represents the detection efficiency.}
\be
\label{event}
R(s) = \sigma(s) \cal{L},
\ee
where $\sigma(s)$ is defined to be the total scattering cross section.
Though the units of cross sections are conventionally  taken as cm$^2$,  
these units are much too big to use for sub-atomic particle scattering, 
and thus more suitable units, called a {\it barn,} are introduced   
$$
1\ {\rm cm}^2=10^{24}\ {\rm barn}=10^{27}\ {\rm mb} = 10^{30}\ \mu{\rm b}=
10^{33}\ {\rm nb}=10^{36}\ {\rm pb}=10^{39}\ {\rm fb}=10^{42}\ {\rm ab.}
$$
It may also be convenient to use these units for the luminosity accordingly
like 
$$ 1\ {\rm cm}^{-2}\ {\rm s}^{-1} =  10^{-33}\ \nbi\ {\rm s}^{-1}.  $$ 
In fact, it is often quite relevant to ask a year long accumulation of the
luminosity, or an integrated luminosity over time. 
It is therefore useful to remember a collider's luminosity
in the units\footnote{Approximately,y
1 year $\approx \pi\times 10^7$ s. It is common that a collider only
operates about $1/\pi$ of the time a year, so it is customary 
to take 1 year $\to 10^7$ s.}
$$10^{33}\ {\rm cm}^{-2}\ {\rm s}^{-1}=1\  \nbi\ {\rm s}^{-1}\approx10\ 
\fbi/{\rm year.}$$

In practice, the instantaneous  luminosity  has some spread around the 
peak energy $\sqrt s$, written as $dL/d\tau$ with $\tau=\hat s/s$ where
$\hat s$ is the c.m.~energy squared with which the reaction actually occurs. 
The more general form for Eq.~(\ref{event}) is
\be
\label{events}
R(s) =  {\cal L} \int d\tau {dL\over d\tau}\ \sigma(\hat s).
\label{lum}
\ee
 With the normalization $\int dL/d\tau\ d\tau=1$, then ${\cal L}$ is the
 peak instantaneous luminosity. The energy spectrum of the luminosity often
can be parameterized by a Gaussian distribution with an energy spread
as given by $\delta E$ ($\sim \sqrt s- \sqrt{\hat s}$) in Tables \ref{eecolliders} and \ref{hcolliders}.
For most of the purposes, the energy spread is much smaller than other energy
scales of interest, so that the luminosity spectrum is well approximated
by  $\delta(1-\tau)$. Thus, Eq.~(\ref{event}) is valid after the proper 
convolution. The only exception would be for  resonant production with a physical
width narrower than the energy spread.  
We will discuss this case briefly  in the next $\epem$ collider section. 

While the luminosity is a machine characteristics, 
the cross section is determined by the fundamental interaction 
properties of the particles in the initial and final states.  Determining the
reaction cross section and studying the scattering properties as a function
of energy, momentum, and angular variables will be of ultimate importance
to uncover new dynamics at higher energy thresholds.

The  electrons and protons are good candidates for serving
as the colliding beams. They are electrically charged so that they can be
accelerated by electric field, and are stable so that they can be put in a 
storage ring for reuse to increase luminosity. 
In Table \ref{eecolliders}, we list the important machine parameters for some
$\epem $ colliders as well as some future machines. 
In Table \ref{hcolliders}, we list the important machine parameters for some
colliders. Electron and proton colliders are complementary 
in many aspects for physics exploration, as we will discuss below. 

\subsection{$e^+e^-$ Colliders}

The collisions between electrons and  positrons  have some major  advantages. 
For instance,
\begin{itemize}
\item The $\epem$ interaction is well understood within the standard model
electroweak theory. The SM processes are predictable without large
uncertainties, and  the total event rates and shapes are easily manageable 
in the collider environments. 
\item The system of an electron and a  positron  has zero charge, zero lepton
number etc., so that it is suitable to create new particles after $\epem$
annihilation.
\item With symmetric beams between the electrons and positrons,  
the laboratory frame is the same as the c.m.~frame, 
so that the total c.m.~energy is fully exploited to reach the highest possible physics
threshold.
\item With well-understood beam properties, the scattering kinematics is 
well-constrained. 
\item It is possible to achieve high degrees of beam polarizations, so that chiral
couplings and other asymmetries can be effectively explored.
\end{itemize}

One disadvantage is the limiting factor due to the large synchrotron radiation
as given in Eq.~(\ref{synch}). The rather light mass of the electrons limits the
available c.m.~energy for an $\epem$ collider. Also, 
 a multi-hundred GeV $\epem$ collider will have to be made a linear 
accelerator \cite{ILC}. This in turn becomes a major 
challenge for achieving a high
luminosity when a storage ring is not utilized.
When performing realistic simulations for high energy $\epem,\ e^-e^-$
reactions at high luminosities, the beamstrahlung effects on the
luminosity and the c.m.~energy  become
substantial and should not be overlooked. 

Another disadvantage for $\epem$ collisions is that they predominantly couple
to a vector (spin 1) state in $s$-channel, so that the resonant production
of a spin-0 state (Higgs-like) is highly suppressed.  For a higher spin state,
such as spin-2, the resonant production will have to go through a higher
partial wave. 

\begin{figure}[t,b]
\psfig{figure=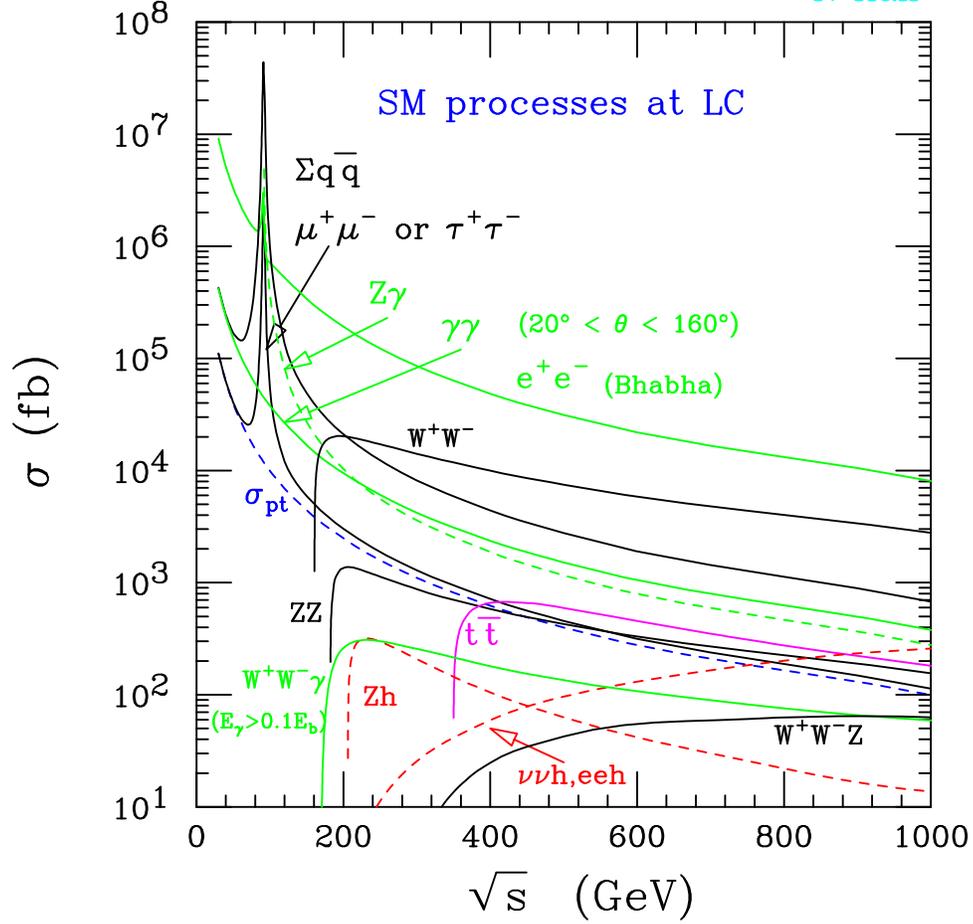,height=5in}
\caption{Scattering cross sections versus c.m.~energy for the SM processes in $e^+e^-$
collisioins. The Higgs boson mass has been taken as 120 GeV.}
\label{fig:epem}
\end{figure}

\subsubsection{Production cross sections for standard model processes}
 
For the production of  two-particle $a,b$ and for unpolarized beams so that the
azimuthal angle can be trivially integrated out  (see Appendix \ref{two-body}),
the differential cross section as a function of the scattering polar angle in the
c.m.~frame is given by
\bea
{d\sigma(e^+e^-\to ab) \over d\cos\theta} = \frac{\beta}{32\pi s}  
\overline{ \sum} |\cM|^2 ,
\label{dsdc}
\eea
where $\beta=\lambda^{1/2}(1, m_a^2/s, m_b^2/s)$ is the speed factor
for the out-going particles,  $\overline{\sum}|\cM|^2$ is 
the scattering matrix element squared, 
summed and averaged over unobserved quantum numbers
like color and spins. 

It is quite common that one needs to consider a fermion pair production
$e^-e^+\to f\bar f$. For most of the situations, the scattering matrix element
can be casted into a $V\pm A$ chiral structure of the form (sometimes with
the help of Fierz transformations)
\bea
\cM = {e^2 \over s} Q_{\alpha\beta}\ 
[\bar v_{e^+}(p_2) \gamma^\mu P_\alpha u_{e^-}(p_1)]\ 
   [\bar \psi_f(q_1) \gamma_\mu P_\beta \psi'_{\bar f}(q_2)],
\label{me}
\eea
where $\alpha, \beta=L, R$ are the  chiral indices, $P_{L,R}^{}=(1\mp \gamma_5)/2$, 
and $Q_{\alpha\beta}$ are the chiral bilinear couplings governed by the underlying
physics of the interactions with the intermediate propagating fields. With this
structure, the  scattering matrix element squared can be conveniently
expressed as
\bea 
\nonumber
\overline{ \sum} |\cM|^2 &=& {e^4 \over s^2} \left[
( |Q_{LL}|^2 + |Q_{RR}|^2 )\ u_i u_j +
( |Q_{LR}|^2 + |Q_{RL}|^2 )\ t_i t_j  \right.  \\
&+& \left. 2Re ( Q_{LL}^* Q_{LR} + Q_{RR}^*Q_{RL} ) m_f m_{\bar f} s \right],
\label{sqme}
\eea
where $t_i=t-m_i^2=(p_1-q_1)^2-m_i^2$ and $u_i=u-m_i^2=(p_1-q_2)^2-m_i^2$.

{
\vskip 0.2cm
\noindent
\tt Exercise: Derive Eq.~(\ref{sqme}) by explicit calculations
from Eq.~(\ref{me}).
\vskip 0.2cm
}

Figure \ref{fig:epem} shows the cross sections for various SM processes
in $\epem$ collisions versus the c.m.~energies. 
The simplest reaction  is the QED process 
$e^+e^-\to\gamma^*\to \mu^+\mu^-$ and its cross section is given by
\bea
{\sigma(e^+e^-\to \gamma^*\to \mu^+\mu^-)\equiv \sigma_{pt}= \frac{4\pi\alpha^2}{3 s}  }.
\eea
In fact, $\sigma_{pt}\approx 100$ fb$/(\sqrt s/{\rm TeV})^2$ 
has become  standard units to measure the size of cross sections
in $e^+e^-$ collisions. However, at energies near the EW scale,  the SM $Z$ boson resonant
production dominates the cross section, seen  as the sharp peek 
slightly below 100 GeV. Above
the resonance, cross sections scale asymptotically as $1/s$, like the 
$s$-channel processes typically do. This is even true for scattering 
with $t,u$-channel diagrams at finite scattering angles. 
The only exceptions are the processes induced by collinear radiations of gauge
bosons off fermions, 
where the total cross section receives a logarithmic enhancement over the 
fermion energy.  For a massive gauge boson fusion process,
\be
\sigma \sim {1\over M^2_V}\ln^2{s\over M_V^2}.
\ee
To have a quantitative feeling, we should know the sizes of typical cross sections
at a 500 GeV ILC
\bea
\nonumber
&&\sigma(W^+W^-) \approx 20\sigma_{pt} \approx 8\ \pb;\\
\nonumber
&&\sigma(ZZ) \approx \sigma(t\bar t) \approx \sigma_{pt} \approx 400\ \fb;\\
\nonumber
&&\sigma(ZH) \approx \sigma(WW\to H) \approx \sigma_{pt}/4 \approx 100\ \fb;\\
\nonumber
&& \sigma(WWZ) \approx 0.1\ \sigma_{pt} \approx 40\ \fb.
\eea

Now let us treat these two important cases in more details.

\subsubsection{Resonant production}
The resonant production for a single particle of mass $M_V$, total width $\Gamma_V$, 
and spin $j$ at c.m.~energy $\sqrt s$ is 
\bea
\sigma(e^+e^-\to V \to X) 
= \frac{4\pi(2j+1)\Gamma(V\to \epem)\Gamma(V\to X)} 
{(s-M^2_V)^2 + \Gamma_V^2 M_V^2} \ {s\over M_V^2},
\label{reson}
\eea 
where $\Gamma(V\to \epem)$ and $\Gamma(V\to X)$ are the partial decay widths
for $V$ to decay to the initial and final sates, respectively. 
This is the Breit-Wigner resonance to be discussed in Eq.~(\ref{BW}) of 
Appendix B. 
For an ideal monochromatic luminosity spectrum,
or the energy spread of the machine much smaller than the physical width $\Gamma_V$,
the above equation is valid.  This  is how the $Z$ 
resonant production cross section was calculated in Fig.~\ref{fig:epem}
as a function of the c.m.~energy $\sqrt s$, and
how the $Z$ line-shape was measured by the energy-scan 
in the LEP I and SLC experiments. 

{
\vskip 0.2cm
\noindent
\tt Exercise: Verify Eq.~(\ref{reson}) by assuming a generic vector production and decay.
\vskip 0.2cm
}

It can occur that the energy spectrum of the luminosity is broader than the narrow
resonant width. One could take the narrow-width approximation as given in
Eq.~(\ref{BW}) and thus the cross section is 
\bea
\sigma(e^+e^-\to V \to X) 
= \frac{4\pi^2 (2j+1)\Gamma(V\to \epem) BF(V\to X)} { M_V^3} \ 
{dL\over d\tau}|_{s=M_V^2},
\label{narrow}
\eea 
where $dL/ d\tau|_{s=M_V^2}$ presents the  contribution
of the luminosity at the resonant mass region . 
In other complicated cases when neither approximation applies between
$\delta E$ and $\Gamma_V$, the more general convolution 
as in Eq.~(\ref{events})
may be needed. For a discussion, see {\it e.g.} the reference \cite{muon}.

{
\vskip 0.2cm
\noindent
\tt Exercise: Derive Eq.~(\ref{narrow}) by applying Eq.~(\ref{lum}) with the narrow 
 width approximation as in Eq.~(\ref{BW}).
\vskip 0.2cm
}

The resonant production is related to the $s$-channel singularity in the 
$S$-matrix for an on-shell  particle propagation. 
It is the most important mechanism for discovering
new particles in high energy collider experiments. We will explore in
great detail the kinematical features in Sec.~\ref{uncover}.

\subsubsection{Effective photon approximation}
\label{WW}

\begin{center}
\begin{figure}[tb]
\hskip 1.2cm
\psfig{figure=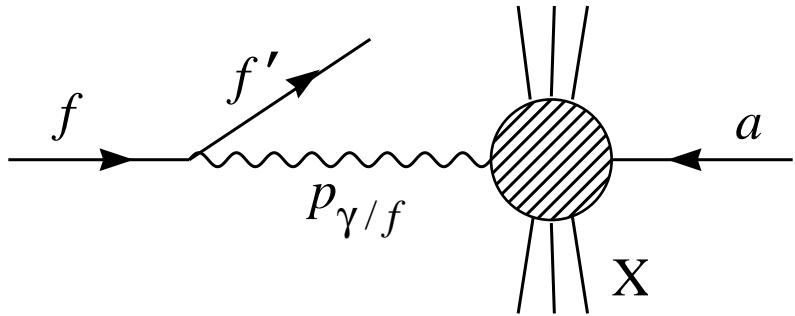,height=1.8in} 
\caption{Illustrative Feynman diagram for effective photon approximation.
\label{fig:ea}}
\end{figure}
\end{center}

A qualitatively different process is initiated from gauge boson radiation,
typically off fermions.  
The simplest case is the photon radiation off an electron. For an electron
of energy $E$, the probability of finding a collinear photon of energy $xE$
is given by 
\bea
P_{\gamma/e}(x)  = \frac{\alpha}{ 2\pi}{ 1+(1-x)^2\over  x} \ln{E^2\over m^2_e},
\label{eq:WW}
\eea
which is known as the Weizs\"acker-Williams spectrum. We see that the
electron mass enters the log to regularize the collinear singularity and $1/x$
leads to the infrared behavior of the photon. 
These dominant features  are a result of a $t$-channel singularity
for the photon. 
This distribution can be obtained by calculating the splitting process
as depicted in Fig.~\ref{fig:ea}, for $e^- a\to e^- X$. 
The dominant contribution is induced by the collinear photon
and thus can be expressed as
\bea
\sigma(e^-a\to e^-X) \approx \int dx\  P_{\gamma/e}(x)\  \sigma(\gamma a\to X).
\label{epa}
\eea
This is also called the effective photon approximation.

This picture of the photon probability distribution  in Eq.~(\ref{epa}) 
essentially treats the photons as initial state to induce the reaction. 
It is also valid for other photon spectrum.
It has been proposed recently to produce much harder photon spectrum
based on the back-scattering laser techniques \cite{gamma} to
construct a ``photon collider". There have been dedicated workshops to
study the physics opportunities for $\epem$ linear colliders operating
in such a photon collider mode, but we will not discuss the details further here. 

A similar picture may be envisioned for the radiation of massive gauge
bosons off the energetic fermions, for example the electroweak gauge bosons
$V=W^\pm,Z$. This is often called the Effective $W$-Approximation \cite{sally,gordy}.
Although the collinear radiation would not be
a good approximation until reaching very high energies $\sqrt s\gg M_V$,
it is instructive to consider the qualitative features, which we will defer 
to Sec.~\ref{EWA} for detailed discussions.

\subsubsection{Beam polarization}
One of the merits for an $\epem$ linear collider is the possible high
polarization for both beams, as indicated in Table \ref{eecolliders}.
Consider first  the  longitudinal polarization along the beam line
direction.  Denote the average $e^\pm$ beam polarization by $P^L_\pm$, with 
$P^L_\pm=-1$ purely left-handed and  $+1$ purely right-handed. Then the
polarized squared matrix element can be constructed \cite{polar}
based on the helicity amplitudes $\cM_{\sigma_{e-} \sigma_{e+} }$
\bea
\nonumber
\overline{\sum} |\cM|^2 = {1\over 4}[
(1-P^L_-)(1-P^L_+) |\cM_{- -} |^2 +
(1-P^L_-)(1+P^L_+) |\cM_{- +}|^2 \\
+ (1+P^L_-(1-P^L_+) |\cM_{+ -}|^2 +
(1+P^L_-)(1+P^L_+) |\cM_{+ +}|^2 ].
\label{pola}
\eea
Since the electroweak interactions of the SM and beyond are chiral,
it is important to notice that contributions from certain helicity
amplitudes can be suppressed or enhanced by properly choosing
the beam polarizations. Furthermore, it is even possible to produce
transversely polarized beams with the help of a spin-rotator. If the 
beams present  average polarizations with respect to a specific 
direction perpendicular to the beam line direction, $-1 < P_\pm^T < 1$,
then there will be one additional term in Eq.~(\ref{pola})
(in the limit  $m_e\to 0$),
\bea
\nonumber
{1\over 4}\ 2\ P^T_- P^T_+\ {\rm Re}( \cM_{-+}\cM_{+-}^*).
\eea
The transverse polarization is particularly important when the
interactions under consideration produce an asymmetry in
azimuthal angle, such as the effect of CP violation.

For a comprehensive count on physics potential for the
beam polarization, see a recent study in Ref.~\cite{epolar}.

\begin{center}
\begin{figure}[tb]
\psfig{figure=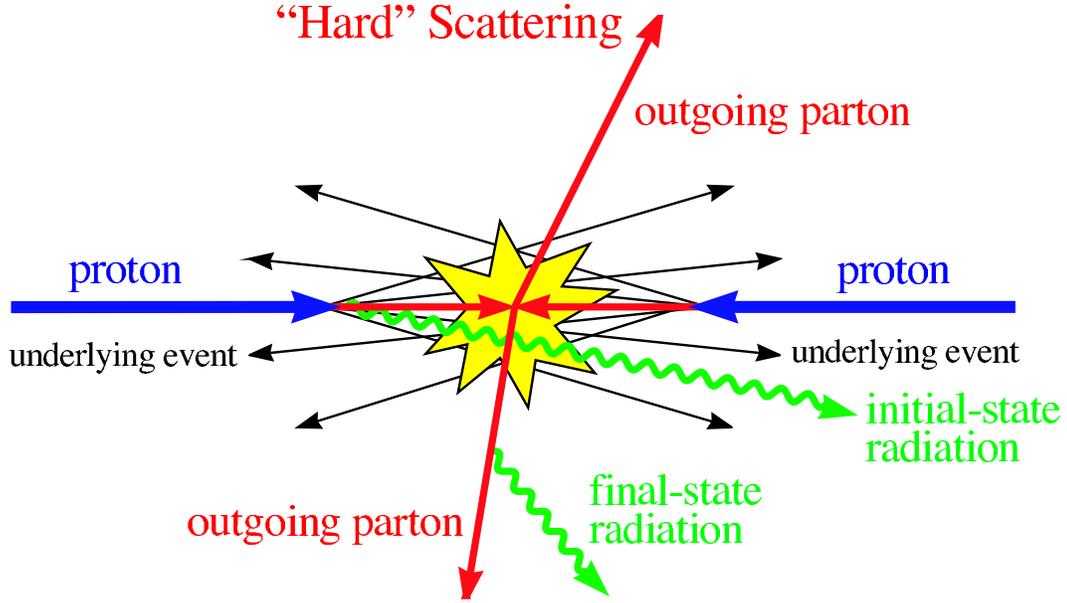,height=3.2in}
\caption{An illustrative event in hadronic collisions.
\label{fig:event}}
\end{figure}
\end{center}

\subsection{Hadron Colliders}
Protons are composite particles, made of  ``partons" of quark and gluons. 
The quarks and gluons are the fundamental degrees of freedom to
participate in strong reactions at high energies according to QCD \cite{george}.
The proton is much heavier than the electron. 
These lead to important differences  between  a hadron collider and 
an $\epem$ collider.
\begin{itemize}
\item Due to the heavier mass of the proton,   hadron colliders can provide 
much higher c.m.~energies in head-on collisions. 
\item Higher luminosity can be achieved also, by making use of the
storage ring for recycle of protons and antiprotons. 
\item Protons participate in strong interactions and thus hadronic
reactions yield large cross sections. The total cross section for a proton-proton
 scattering can be estimated by dimensional analysis to be about 100 mb,
 with weak energy-dependence.
\item
At higher energies, there are many possible channels open up 
resonant productions for
 different charge and spin states,  induced by the initial parton 
 combinations  such as  $q\bar q$, $qg$, and $gg$.
As discussed in the last section for gauge boson radiation, there are also
contributions like initial state $WW, ZZ$ and $WZ$ fusion.
\end{itemize}

The compositeness and the strong interactions of the protons 
on the other hand can  be disadvantageous  in certain aspect,
as we will see soon. An interesting event for a high-energy hadronic  
scattering may be illustrated by Fig.~\ref{fig:event}.

\subsubsection{Hard scattering of partons}
Thanks to the QCD factorization theorem,  which states that 
the cross sections for  high energy hadronic reactions with a large
momentum transfer can be factorized into a parton-level ``hard scattering"
convoluted with the parton ``distribution functions". 
For scattering of two hadrons $A$ and $B$ to produce a final state
$F$ of our interest, 
the cross section can be  formally written  as a sum over
the sub-process cross sections from the contributing partons
\bea
\sigma(AB\to F\ X) = \sum_{a,b} \int dx_1 dx_2\  P_{a/A}(x_1,Q^2)  P_{b/B}(x_2,Q^2)
\  \hat\sigma(ab \to F),
\label{eq:ab}
\eea
where $X$ is the inclusive scattering remnant, and 
$Q^2$ is the factorization scale (or the typical momentum transfer) 
in the hard scattering process, much  larger than 
$\Lambda_{QCD}^2\approx (200\  \mev)^2$.
The parton-level hard scattering cross section can be calculated 
perturbatively in QCD, while the parton distribution  functions 
parameterize the non-perturbative aspect and can be only
obtained by some ansatz and by ffitting the data. For more discussions,
the readers are referred to George Sterman's lectures \cite{george} on
QCD,  or the excellent text \cite{esw} on thesse topics.

\begin{center}
\begin{figure}[tb]
\psfig{figure=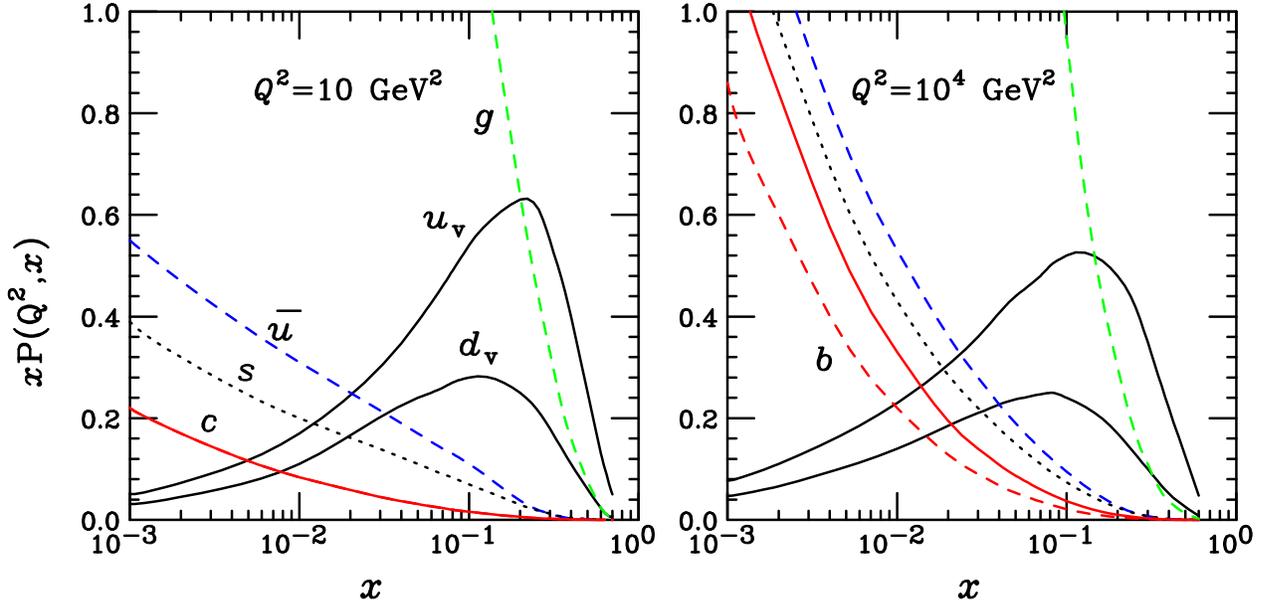,height=3.2in}
\caption{Parton momentum distributions versus their energy fraction $x$ at two
different factorization scales, from CTEQ-5. }
\label{fig:pdfs}
\end{figure}
\end{center}
 
Since the QCD parton model plays a pivotal role in understanding
hadron collisions and uncovering new phenomena at high energies,
we plot in Fig.~\ref{fig:pdfs} the parton momentum distributions 
versus the energy fractions $x$, taking CTEQ-5 as a 
representative \cite{cteq5}. 
For comparison, we have chosen the QCD factorization scale to be 
$Q^2$=10 $\gev^2$ and $10^4\ \gev^2$ in these two panels,
respectively.  Several general features are important
to note for future discussions. The valence quarks $u_v,\ d_v$,
as well as the gluons carry a large momentum fraction, typically
$x\sim 0.08 - 0.3$. The ``sea quarks" ($\bar u=u_{sea},
\bar d=d_{sea}, s,c,b$) have
small $x$, and are significantly enhanced at higher $Q^2$. Both
of these features lead to important collider consequences. First of
all, heavy objects near the energy threshold are more likely produced
via valence quarks. Second, higher energy processes (comparing to
the mass scale of the parton-level subprocess) are more dominantly
mediated via sea quarks and gluons.

\begin{center}
\begin{figure}[tb]
\psfig{figure=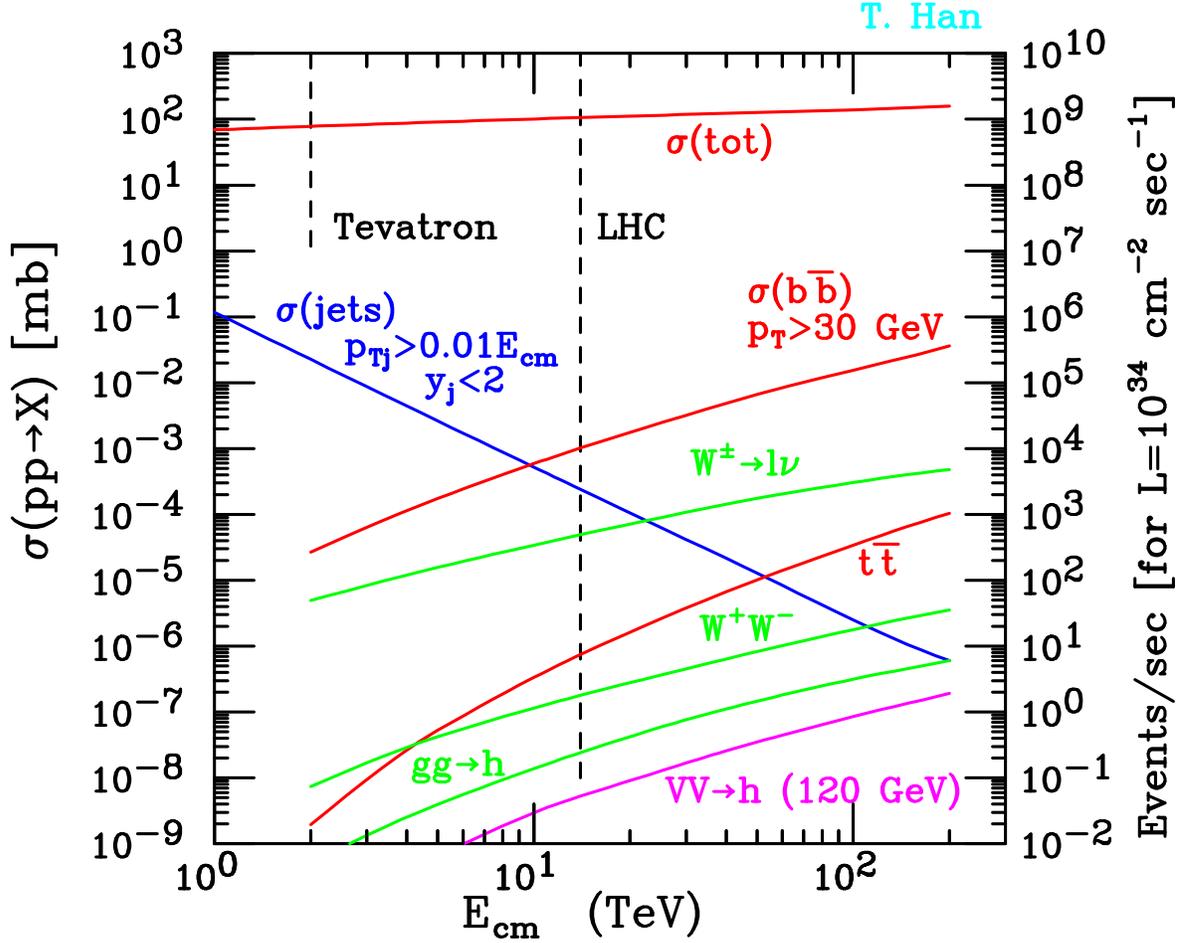,height=5in}
\caption{Scattering cross sections versus c.m.~energy
for the SM processes in $pp$ collisioins. 
 The Higgs boson mass has been taken as 120 GeV.}
\label{fig:hadron}
\end{figure}
\end{center}

\subsubsection{Production cross sections for standard model processes} 
In Figure \ref{fig:hadron}, we show the integrated cross sections for various typical 
processes in the SM versus c.m.~energy of a $pp$ hadron collider in units of mb.  
The scale on the right-hand side gives the event rate for an instantaneous luminosity
$10^{34}\ {\rm cm}^{-2}\ {\rm s}^{-1}$, a canonical value for the LHC.
We have indicated the energies at the Tevatron 
and the LHC by the vertical dashed lines. 
First of all, we have plotted the $pp$ total cross section as the line on 
the top. It is known that the cross section increases with the 
c.m.~energy \cite{ttwu}.
Unitarity argument implies that it can only increase as a power of $\ln s$. 
An empirical scaling relation $s^{0.09}$ gives a good fit to the 
measurements upto date, and has been used here. 

All integrated cross sections in hadronic collisions increase with the 
c.m.~energy due to the larger parton densities at higher energies. 
The jet-inclusive cross section $\sigma$(jets) is given by the blue line.
The reason the cross section falls is due to our choice of an
energy-dependent cut on the jet's transverse momuntum. 
The $b\bar b$ pair production is also
sensitively dependent upon the transverse momentum cutoff since the
mass $m_b$ is vanishingly small comparing to the collider energies
and thus the integrated cross section presents the familiar collinear 
singularity in the forward scattering region. 
The production at the leading order is dominantly via the gluon-initiated
process $gg\to b \bar b$, and is of the order of 1 $\mu$b at the LHC
energy (with a cutoff $p_T>30$ GeV). The top-quark production is
again dominated by the gluon fusion, leading to about $90\%$ of the
total events. The rate of the leading order prediction 
is about 700 pb or about 7 Hz with a canonical luminosity, and
higher order QCD corrections are known to be substantial \cite{top}.
It is thus justifiable to call the LHC a ``top-quark factory".
We also see that the leading Higgs boson production mechanism 
is also via the gluon fusion, yielding about 30 pb. QCD corrections
again are very large, increasing the LO cross section by a significant
factor \cite{ggh}. Another interesting
production channel is the gauge-boson fusion $VV\to h$, that is
about factor of 5 smaller than the inclusive $gg\to h$ in rate,
and the QCD correction is very modest \cite{vvh}. 

Most of the particles produced in high-energy collisions are unstable.
One would need very sophisticated modern detector complex and
electronic system to record the events for further analyses. 
We now briefly discuss the basic components for collider detectors.

\section{Collider Detectors:  Our Electronic Eyes}
\label{detector}

Accelerators and colliders are our powerful tools to produce 
scattering events at high energies. Detectors are our  ``e-eyes"
to record and identify the useful events to reveal the nature of 
fundamental interactions.

\subsection{Particle Detector at Colliders}
The particle detection is based on its interactions with matter of which
the detectors are made. 
A modern particle detector is an electronic complex beyond the traditional
particle detection techniques,  which typically consists of a 
secondary displaced vertex detector/charge-tracking system, 
electromagnetic calorimetry, hadronic calorimetry and a muon chamber, etc. 

\begin{center}
\begin{figure}[tb]
\psfig{figure=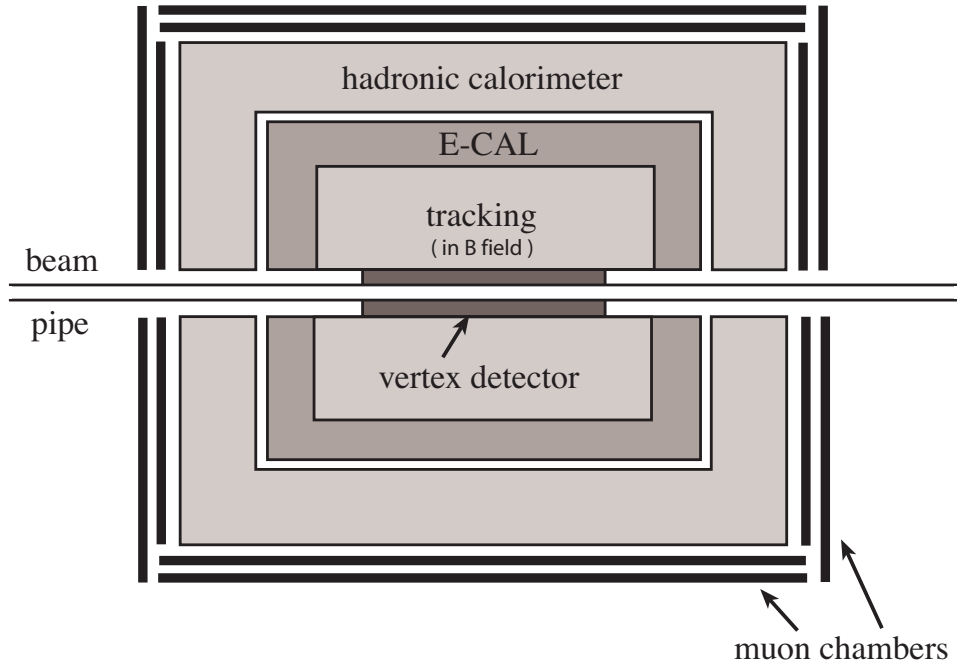,height=3.5in}
\caption{Modern multi-purpose detector at colliders.
\label{fig:detect}}
\end{figure}
\end{center}

A simplified layout is shown in Fig.~\ref{fig:detect}.

\subsection{What Do Particles Look Like in a Detector}

\begin{center}
\begin{figure}[tb]
\psfig{figure=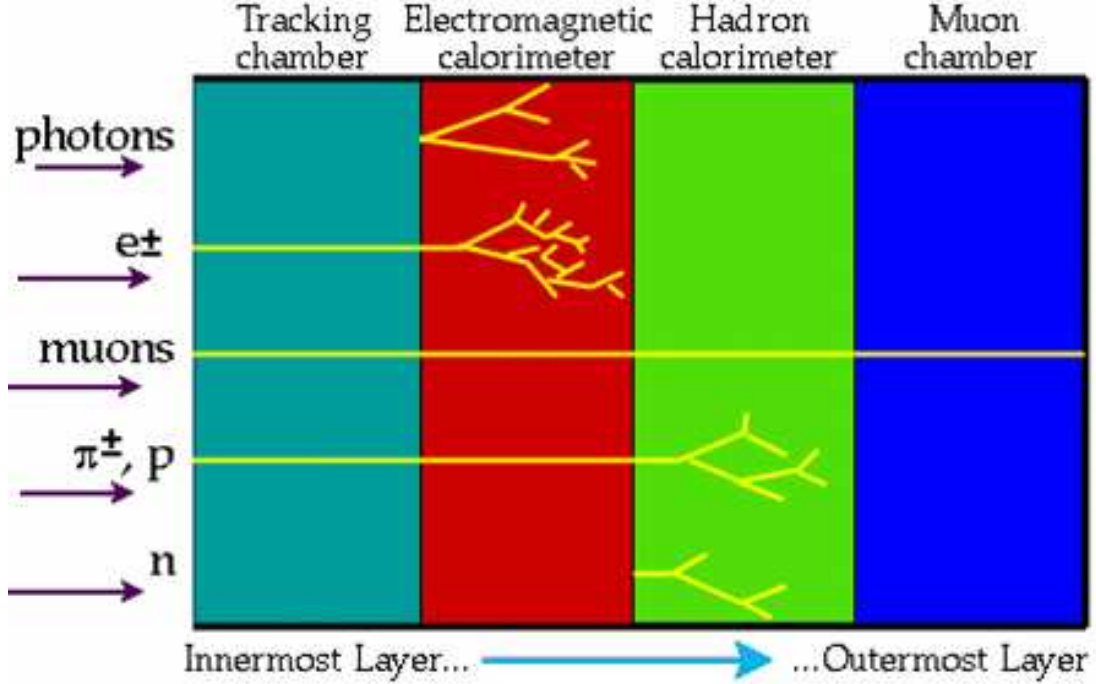,height=3.6in}
\caption{Particle signatures left in the detector components.
\label{fig:showup}}
\end{figure}
\end{center}

As theorists, we  mostly deal with the fundamental degrees of freedom 
in our SM Lagrangian, 
namely the quarks, leptons, gauge bosons etc. in our calculations.
 The truth is that most of them  are not  the particles
directly ``seen" in the detectors.  Heavy particles like $Z,\ W,\ t$ 
will promptly decay to leptons and quarks, with a lifetime
$1/\Gamma \sim 1/(2\ \gev) \approx 3.3\times 10^{-25}$ s.
Other quarks will fragment into color-singlet hadrons due to 
QCD confinement at a time scale of 
$t_h\sim 1/\Lambda_{QCD} \approx 1/(200\ \mev)\approx 3.3\times10^{-24}$ s.
The individual hadrons from fragmentation may even behave
rather differently in the detector, depending on their interactions with
matter and their life times. 
Stable paricles such as $p,\ \bar p,\ e^\pm,\ \gamma$  will
show up in the detector as energy deposit in hadronic and electromagnetic
calorimeters or charge tracks in the tracking system.
In Fig.~\ref{fig:showup},  we indicate what particles may leave 
what signatures in certain components of the detector.

\begin{center}
\begin{figure}[tb]
\psfig{figure=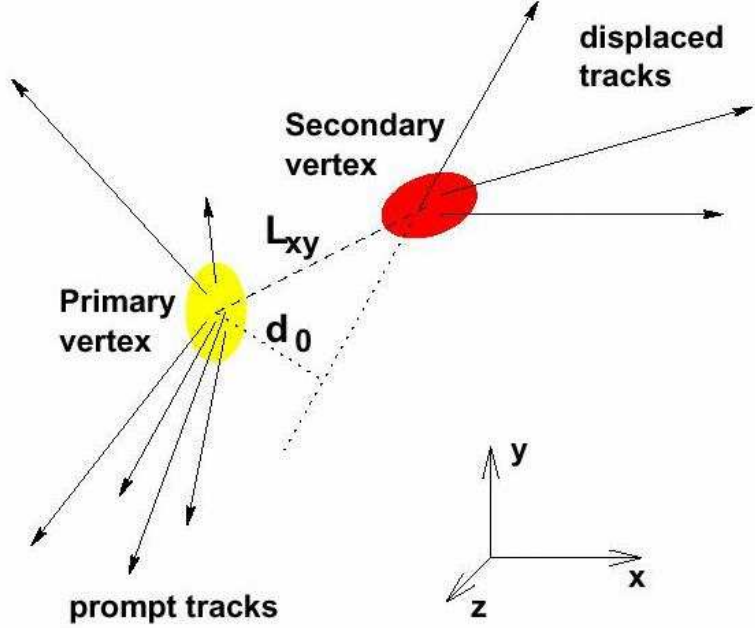,height=3.4in}
\caption{An illustrative event leading to a displaced secondary vertex.
\label{fig:vertex}}
\end{figure}
\end{center}

In order to have  better understanding for the particle observation, 
let us recall the decay length of an unstable particle
\begin{equation}
d=(\beta\ c \tau)\gamma \approx (300\ \mu m) ({\tau\over 10^{-12}\ s})\ \gamma,
\end{equation}
where $\tau$ is the particle's proper  lifetime and $\gamma=E/m$ is the relativistic factor. 
We now can comment on how particles may show up in a detector.
\begin{itemize}
\item
Quasi-stable:  fast-moving particles of a life-time $\tau > 10^{-10}$ s
will still interact in the detector in a similar way. 
Those include the weak-decay particles like the neutral hadrons 
$n,\Lambda,K^0_L,\ ... $ and charged particles $ \mu^\pm,\ \pi^\pm, K^\pm,\ ...$
\item Short-lived resonances: particles undergoing a decay of  typical  
electromagnetic or strong strength, such as $\pi^{0},\ \rho^{0,\pm}$... and
very massive particles like 
$Z,W^\pm,t, (H ...)$, will decay ``instantaneously". They can be only ``seen"
from their decay products and hopefully via a reconstructed resonance.
\item
displaced vertex:  particles of a life-time $\tau \sim 10^{-12}$ s, such as
$B^{0,\pm},\ D^{0,\pm},\ \tau^\pm,$ may travel a distinguishable distance
($c\tau \sim 100\ \mu$m.)
before decaying into charged tracks, and thus result in a displaced
secondary vertex, as shown in Fig.~\ref{fig:vertex}, where the
decay length between the two vertices is denoted by $L$.
As an interesting and important case, $K_S^0$ with $c\tau \sim 2.7$ cm
also often results in a secondary vertex via its decay to $\pi^+\pi^-$.
\item
Things not ``seen":  those that do not participate in electromagnetic nor
strong interactions, but long-lived as least like the quasi-stable particles,
will escape from detection by the detector, such as the neutrinos $\nu$
and neutralinos $\tilde\chi^0$ in SUSY theories, etc.
\end{itemize}

Now coming back to the elementary particles in the SM, we 
illustrate their behavior  in Table \ref{detectsm}. 
A check indicates an appearance in that component, a cross
means no, a check-cross is partially yes. Other symbols are self-explained.

\begin{table}[htb]
\begin{tabular}{|c|c c c c c |} 
\hline
Leptons  & Vetexing   & Tracking &  ECAL &  HCAL & Muon Cham. \\
\hline$e^\pm$ & $\times$ & $\vec p$ &   $E$   &   $\times$        & $\times$      \\
$\mu^\pm$  &  $\times$ &  $\vec p$   &   $\surd$    &  $\surd$   & $\vec p$ \\
$\tau^\pm$  & $\surd\times$  & $\surd $   &   $e^\pm$    &  $h^\pm;\ 3h^\pm$   & $\mu^\pm$ \\
$\nu_e,\nu_\mu,\nu_\tau$ & $\times$  & $\times$      &   $\times$   &   $\times$    & $\times$     \\
\hline
Quarks  &  & & & & \\
\hline
$u,d,s$ &  $\times$ & $\surd$  & $\surd$  & $\surd$     & $\times$      \\
$c\to D$    &  $\surd$ &   $\surd$ &   $e^\pm$    &  $h$'s  & $\mu^\pm$ \\
$b\to B$   &  $\surd$  & $\surd$   &   $e^\pm$    &  $h$'s  & $\mu^\pm$ \\
$t\to bW^\pm$ &  $b$  &  $\surd$  &   $e^\pm$    &  $b+2$ jets  & $\mu^\pm$ \\
\hline
Gauge bosons  &  & & & & \\
\hline
$\gamma$ & $\times$  & $\times$ &   $E$   &   $\times$        & $\times$      \\
$g$  &  $\times$ &  $\surd$  &   $\surd$    &  $\surd$   & $\times$ \\
$W^\pm\to \ell^\pm \nu$ &  $\times$ & $\vec p$  &   $e^\pm$    &  $\times$  & $\mu^\pm$ \\
$\quad~ \to q\bar q'$ &  $\times$ &  $\surd$ &   $\surd$    &  2 jets  & $\times$ \\
$ Z^0\to \ell^+\ell^-$ &  $\times$  & $\vec p$  &   $e^\pm$    &  $\times$ & $\mu^\pm$ \\
$ \to q\bar q$ &  $(b\bar b)$  & $\surd$  &   $\surd$    &  2 jets  & $\times$ \\
\hline
\end{tabular}
\caption{What the elementary particles in the SM look like in detectors.}
\label{detectsm} 
\end{table}

\subsection{More on Measurements}

It is  informative to discuss in a bit more detail a few main
components for the particle detection. We hope to indicate how and 
how well the energy, momentum, and other properties of particles
can be measured. When needed, we will freely take either the ATLAS 
or the CMS detector as an example  for the purpose of illustration.

\vskip 0.3cm
\noindent
\underline{Vertexing:} 
Normally, at least two charged tracks are needed 
to reconstruct a secondary decay vertex, as illustrated in Fig.~\ref{fig:vertex}.
Note that if the decaying particle moves too fast, then the decay products
will be collimated, with a typical angle $\theta \approx \gamma^{-1} =m/E$.
The impact parameter $d_0$ as in Fig.~\ref{fig:vertex} can be approximated
as $d_0\approx L_{xy} \theta$. 
The impact parameter is crucial to determine the displaced vertex.
For instance, the ATLAS detector \cite{ATLAS} has the resolution 
parameterized by 
\be
{\Delta d_0} = 11 \oplus {73\over (\pt/\gev)\ \sqrt{\sin\theta}}\ (\mu m),
\ee
where the notation $\oplus$ implies a sum in quadrature. 

It is possible to resolve a secondary vertex  along the longitudinal direction
alone, which is particularly important if there will be only one charged
track  observed. In this case, the resolution is typically worse and it can
be approximated \cite{ATLAS} as
\be
{\Delta z_0} = 87 \oplus {115 \over (\pt/\gev)\  \sqrt{\sin^3\theta}}\ (\mu m).
\ee

\vskip 0.3cm
\noindent
\underline{Tracking:} Tracking chamber determines the trajectories
of traversing charged particles as well as their electromagnetic energy
loss $dE/dx$. The rapidity coverage is 
\be
|\eta_\mu|\approx 2.5,
\ee
for both ATLAS and CMS.
 
When combined with a magnetic field (2 T for ATLAS and 4 T for CMS), 
the system can be used to measure a charged particle momentum.
The curvature of the trajectory is inversely proportional to the particle
momentum
\be
\kappa \equiv {1\over \rho} \propto {QB \over p},
\label{curv}
\ee
where $Q$ is the particle's electric charge and $B$ the external magnetic
field. Therefore, knowing $B$ and assmuing a (unity) charge, the momentum
$p$ can be determined. 

The energy-loss measurement $dE/dx$ for heavy charged particles 
may be used for particle identification. 
For instance, the Bethe-Bloch formula for the
energy loss by excitation and ionization gives a  scaling quadratically 
with the particle charge and inversely with the speed
\be
{dE\over dx} \propto \left({Q \over \beta} \right)^2,
\ee
independent of the charged particle mass. 
The mass can thus be deduced from $p$ and $\beta$. However, if we allow
a most general case for a particle of arbitrary $m,Q$, then an additional 
measurement (such as $\beta$ from a Cerenkov counter or time-of-flight
measurements) would be needed to fully determine the particle identiy.

From the relation Eq.~(\ref{curv}), the momentum resolution based on a
curvature measurement can be generically expressed as
\be
{\Delta \pt \over \pt} = {a \pt} \oplus {b},
\ee
For instance, the ATLAS \cite{ATLAS} (CMS \cite{CMS}) detector 
has the resolution 
parameterized by $a=36\%\ \tev^{-1}\ \ (15\%\ \tev^{-1}),\ 
b=1.3\%/\sqrt{\sin\theta}\ \ (0.5\%)$. In particular, the momentum resolution
for very high energy muons about $\pt \approx 1$ TeV in the central region 
can reach $10\%\ (6\%)$ for ATLAS (CMS). Good curvature 
resolution for highly energetic particle's tracks is important for the 
charge determination.

\vskip 0.3cm
\noindent
\underline{ECAL:} High-energy electrons and photons often lead to
dramatic cascade electromagnetic showers due to bremsstrahlung and
pair production.  The number of particles created increase exponentially
with the depth of the medium.  Since the incident energy to be measured 
by the electromagnetic calorimetry (ECAL) is
proportional to the maximum number of particles created, the energy resolution
is characterized by $1/\sqrt N$, often parameterized by
\be
{\Delta E\over E} = {a\over \sqrt{E/ \gev}} \oplus {b},
\ee
where $a$ is determined by the Gaussian error and $b$ the response
for cracks.
For ATLAS (CMS), $a=10\%\ \ (5\%),\ \ b=0.4\%\ \ (0.55\%)$.

The coverage in the rapidity range can  reach
\be
|\eta_{e,\gamma}|\approx 3
\ee
or slightly over for both ATLAS and CMS.

\vskip 0.3cm
\noindent
\underline{HCAL:} 
Similar to the ECAL, showers of subsequent hadrons  can be developed
from the high-energy incident hadrons. An HCAL is to measure the hadronic
energy, and the Gaussian error again is parameterized as
\be
{\Delta E\over E} = {a\over \sqrt E} \oplus {b}.
\ee
For ATLAS (CMS), $a=80\%\ \ (100\%),\ \ b=15\%\ \ (5\%)$.

The rapidity coverage by the forward hadronic calorimeter can  reach
\be
|\eta_h|\approx 5
\ee
for both ATLAS and CMS.


\vskip 0.2cm
\noindent
\subsection{Triggering}

So far, we have ignored one very important issue: data acquisition and 
triggering.
Consider $pp$ collisions at the LHC energies, the hadronic total
cross section is of the order about 100 mb, 
and the event rate at the designed luminosity ($10^{34}$ cm$^{-2}$ s$^{-1}$)
will be about 1 GHz (compare with the clock speed of your fast 
computer processor). A typical even will take about one Mega bytes
of space. 
It is therefore impossible  for the detector electronic system to record
the complex events of such a high frequency. Furthermore,  the 
physical processes of our interest occur at a rate of $10^{-6}$ lower
or more. Thus, one will have to be very selective in recording events 
of our interest. In contrast, there will be no such problems at $\epem$
colliders due to the much lower reaction rate. 

Trigger is the decision-making process using 
a desired temporal and spatial correlation in the detector signals.
It is provided by examining the properties of the physical process
as appeared in the detector. 
Modern detectors for hadron colliders such
as CDF, D0 at the Tevatron and ATLAS, CMS at the LHC
typically adopt three levels of triggering. 
At the LHC experiments, Level-1 triggering brings
the event rate down to the order of $10^{5}$ Hz;  Level-2  to about
$10^{3}$ Hz;  and Level-3  finally to about 100 Hz to tape. 

There are many means to design a 
trigger, such as particle identification, multiplicity, kinematics, 
and event  topology etc. Modern detectors at colliders usually
can trigger on muons by a muon chamber, 
electrons/photons as electromagnetic objects,
$\tau$/hadrons and jets as hadronic objects, global energy sum
and missing transverse energy, and some combinations of the
above.

%
%
\begin{table}[tb]
\begin{tabular}{|c|c c | } 
\hline
\multicolumn{1}{|c|}{ }&
\multicolumn{2}{c|}{ATLAS} \\
\hline
  Objects & $\eta$   & $\pt$ (GeV)  \\
\hline\hline
$\mu$ inclusive  &  2.4 &  6\ (20)     \\
\hline
$e$/photon inclusive & 2.5 &   17  (26) \\
Two $e$'s or two photons & 2.5 & 12 (15)    \\
\hline
1-jet inclusive  & 3.2  & 180 (290)  \\
3 jets  & 3.2  &    75 (130) \\
4 jets  & 3.2  &   55 (90)  \\
$\tau$/hadrons  & 2.5 & 43\ (65)   \\
\hline
$\etmiss$ & 4.9  & 100   \\
Jets$+\etmiss$ & 3.2,\ 4.9 & 50,50 (100,100)  \\
\hline
\end{tabular}
\caption{Level-1 trigger thresholds in $\eta$-$\pt$ coverage 
for the ATLAS  {\protect\cite{ATLAS}} detector.
Entries are for a luminosity of $10^{33}$ cm$^{-2}$ s$^{-1}$
($10^{34}$ cm$^{-2}$ s$^{-1}$ in parentheses). }
\label{trigger} 
\end{table}

%

Relevant to collider phenomenology is to know 
what particles may be detected in 
what kinematical region usually in $p_T$-$\eta$ coverage,  
as the detector's acceptance for triggering purposes.
Table \ref{trigger} summarizes Level-1 trigger thresholds for
ATLAS for the commonly observed objects and some useful combinations.

Inversely, if we find some trigger designs inadequate for certain physics 
needs, such as a well-motivated new physics signal with
exotic characteristics or unusual kinematics, 
it is the responsibility of our phenomenologists to communicate 
with our experimental colleagues to propose new trigger designs.

\section{ Uncover New Dynamics at Colliders }
\label{uncover}

Instead of summarizing which new physics scenario can be covered
by which collider to what extent, I would like to discussion a few 
examples for observing signals to illustrate the basic techniques
and the use of kinematics. 
The guiding principles are simple: maximally and optimally make use
of the experimentally accessible observables to uncover new particles
and to probe their interactions. In designing the observables, one will
need to concern their theoretical properties, like under Lorentz transformation,
charge and $C,P,T$ discrete symmetries etc., as well as their experimental
feasibility, like particle identification,  detector acceptance and resolutions etc. 
I hope that this serves the purpose to 
stimulate reader's creativity to cleverly exploit kinematics to reveal 
new  dynamics in collider experiments.


\subsection{Kinematics at Hadron Colliders}

In performing parton model calculations for hadronic collisions like in 
Eq.~(\ref{eq:ab}), the partonic c.m.~frame is not the same as the hadronic
c.m.~frame, {e.~g.}~the lab frame for the collider. 
Consider a collision between two hadrons of $A$ and $B$
of four-momenta $P_A=(E_A^{},0,0,p_A^{})$ and $P_B=(E_A^{},0,0,-p_A^{})$ 
in the lab frame. The two partons participating 
the subprocess have momenta $p_1=x_1 P_A^{}$ and $p_2=x_2 P_B^{}$.
The parton system thus moves in the lab frame with a four-momentum 
\bea
P_{cm}=[(x_1+x_2)E_A^{},\ 0,\ 0,\  (x_1-x_2)p_A^{} ] \quad (E_A\approx p_A^{}) ,
\eea
or with a speed $\beta_{cm} = (x_1-x_2) / (x_1+x_2)$, or with a rapidity 
\be
y_{cm} = {1\over 2} \ln {x_1 \over x_2}.
\label{eq:rapcm}
\ee

Denote the total hadronic c.m.~energy by $S=4E_A^2$ and the partonic c.m.~energy
by $s$, we have
\be
s \equiv \tau S,\quad \tau = x_1 x_2 = {s\over S}.
\ee
The parton energy fractions are thus given by
\be
x_{1,2}^{} = \sqrt\tau\ e^{\pm y^{}_{cm}}.
\ee
One always encounters the integration over the energy fractions
as in Eq.~(\ref{eq:ab}). With this variable change, one has
\be
\int_{\tau_0}^1 dx_1 \int_{\tau_0/x_1}^1 dx_2 = \int_{\tau_0}^1 d\tau
 \int_{{1\over 2}\ln\tau}^{-{1\over 2}\ln\tau} dy_{cm}.
\ee
The variable $\tau$ characterizes the (invariant) mass of the reaction, 
with $\tau^{}_0=m^2_{res}/S$ and $m_{res}$ is the threshold for the parton level
final state (sum over the masses in the final state); while $y_{cm}$ specifies the
longitudinal boost of the partonic c.m.~frame with respect to the lab frame.
It turns out that the $\tau - y_{cm}$ variables are
better for numerical evaluations, in particular with a resonance as we will
see in a later section.

Consider a final state particle of momentum $p^\mu = (E,\vec p)$
in the lab frame. 
Since the c.m.~frame of the two colliding partons is {\it a priori} 
undetermined with respect to the lab frame, 
the scattering polar angle  $\theta$ in these two frames is not
a good observable to describe theory and the experiment.
It would be thus more desirable to seek for kinematical variables
that are invariant under unknown longitudinal boosts.

\vskip 0.2cm
\noindent
\underline{Transverse momentum and the azimuthal angle:}\  Since the 
ambiguous motion between the parton c.m.~frame and the 
hadron lab frame is along
the longitudinal beam direction ($\vec z$), variables involving only
the transverse components are invaraint under longitudinal  boosts. It is thus
convenient, in contrast to Eqs.~(\ref{ps1})  and (\ref{ps2}) of Appendix \ref{app-ps} 
in the spherical coordinate,
 to write the phase space element in the cylindrical coordinate as
\be
{d^3\vec p\over E} = d p_x d p_y {d p_z\over E}  =  \pt d\pt d\phi\  {d p_z\over E} ,
\ee
where $\phi$ is the azimuthal angle about the $\vec z$ axis, and 
\be
\pt = \sqrt{p_x^2 + p_y^2} = p \sin\theta
\ee
 is the transverse momentum. It is obvious
that both $\pt$ and $\phi$ are  boost-invariant, so is $dp_z/E$.

\noindent
{\tt Exercise: Prove that $dp_z/E$ is longitudinally boost-invariant.}


\vskip 0.2cm
\noindent
\underline{Rapidity and pseudo-rapidity:}\  The rapidity of  a particle of momentum
$p^\mu$ is defined to be
\be
y = {1\over 2} \ln{E+p_z \over E - p_z }.
\ee
{\tt Exercise: With the introduction of  rapidity $y$, show that a particle 
four-momentum can be rewritten as 
\be
p^\mu=(E_T \cosh y, \pt \sin\phi, \pt \cos \phi, E_T\sinh y),\quad \et=\sqrt{p_T^2+m^2} .
\ee
The phase space element then can be expressed as
\be
{d^3\vec p\over E}  =  \pt d\pt d\phi\  dy =  \et d\et d\phi\  dy.
\ee  }

Consider the rapidity in a boosted frame (say the parton c.m.~frame), and
perform the Lorentz transformation as in Eq.~(\ref{lor}) of 
Appendix \ref{app-ps},
\be
y' = {1\over 2} \ln{E'+p'_z \over E' - p'_z } = 
{1\over 2}\ln{(1-\beta_0) (E+p_z) \over (1+\beta_0)(E - p_z) } = y - y_0.
\label{eq:rap}
\ee
In the massless limit, $E\approx |\vp|$, so that 
\be
y \to  {1\over 2} \ln{1+\cos\theta \over 1 - \cos\theta }= \ln{\cot {\theta \over 2} } \equiv \eta ,
\ee
where $\eta$ is the pseudo-rapidity, which has one-to-one correspondence with
the scattering polar angle $\pi \ge \theta\ge 0$ for $-\infty < \eta < \infty$. 

Since $y$ as well as $\eta$ is additive under longitudinal boosts as seen in 
Eq.~(\ref{eq:rap}), the rapidity difference 
$\Delta y= y_2 - y_1 = y'_2 - y'_1$ is invariant in the two frames. Thus the shape 
of  rapidity distributions $d\sigma/dy$ 
in the two frames would remain the same
if the boost is by a constant velocity. In realistic hadronic collisions, the boost
 of course varies on an event-by-event basis according to Eq.~(\ref{eq:rapcm})
 and the distribution is generally  smeared.

\begin{center}
\begin{figure}[tb]
\psfig{figure=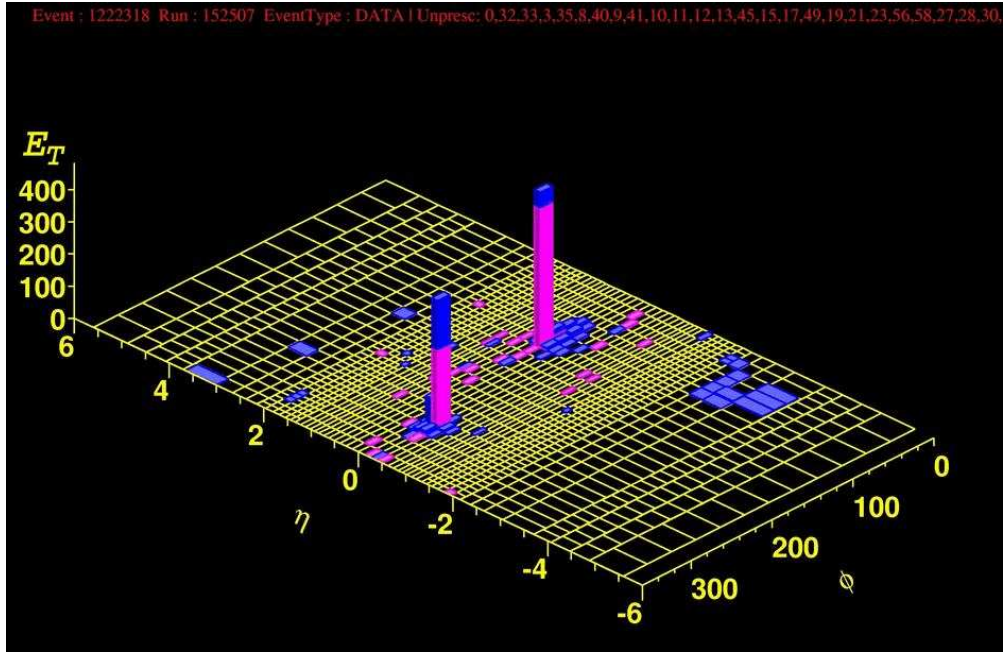,height=3.4in}
\caption{A  CDF di-jet event on a  lego plot in the $\eta-\phi$ plane. 
The height presents the transverse energy scale, and the two colors
(blue and pink) indicate the energy deposit in the two calorimeters
(ECAL and HCAL).
\label{fig:lego}}
\end{figure}
\end{center}

\vskip 0.2cm
\noindent
\underline{The lego plot:} 
It should be clear by now that it is desirable to use the kinematical variables 
$(\pt,\eta,\phi)$ to describe events in hadronic collisions.
In collider experiments, most often, electromagnet and hadronic calorimeters
provide the energy measurements for (essentially) massless particles, such
as $e^\pm,\ \gamma,$ and light quark or gluon jets. Thus
\be
\et = \pt = E\cos\theta = E\cosh^{-1}\eta.
\ee
A commonly adopted presentation for an electromagnetic or hadronic event
is on an  $\eta-\phi$ plane with the height to indicate the transverse energy
deposit $\et$, called the Lego plot. We show one typical  di-jet event from CDF 
by a lego plot in Fig.~\ref{fig:lego}. Of particular importance for a lego plot 
is that the separation between to objects on the plot is invariant under longitudinal 
boosts. This is seen from the definition of separation
\be
\Delta R= \sqrt{ \Delta\eta^2 + \Delta\phi^2}.
\ee
As a quantitative illustration, for two objects back-to-back in the central region, 
typically $\Delta\eta < \Delta\phi$ and $\Delta R \approx \Delta\phi\sim \pi$.

Another important consequence for the introduction of separation is that it
provides a practical definition of a hadronic jet, and $\Delta R$ specifies the 
cone size of a jet formed by multiple hadrons within $\Delta R$.

\subsection{$s$-channel Singularity: Resonance Signals}

\subsubsection{The invariant mass variable}

Searching for a resonant signal in the $s$-channel has been the most
effective way of discovering new particles. Consider an unstable particle
$V$ produced by $a+b$ and decaying to $1+2+...+n$. For a weakly coupled
particle $\Gamma_V \ll M_V$, according to the Breit-Wigner resonance
Eq.~(\ref{BW}), the amplitude develops a kinematical peak near
the pole mass value at 
\be
(p_a+p_b)^2 = (\ \sum_i^np_i \ )^2 \approx M_V^2.
\ee
This is called the invariant mass, and is the most effective observable 
for discovering a resonance if either the initial momenta or the final
momenta can be fully reconstructed. 

\begin{center}
\begin{figure}[tb]
\psfig{figure=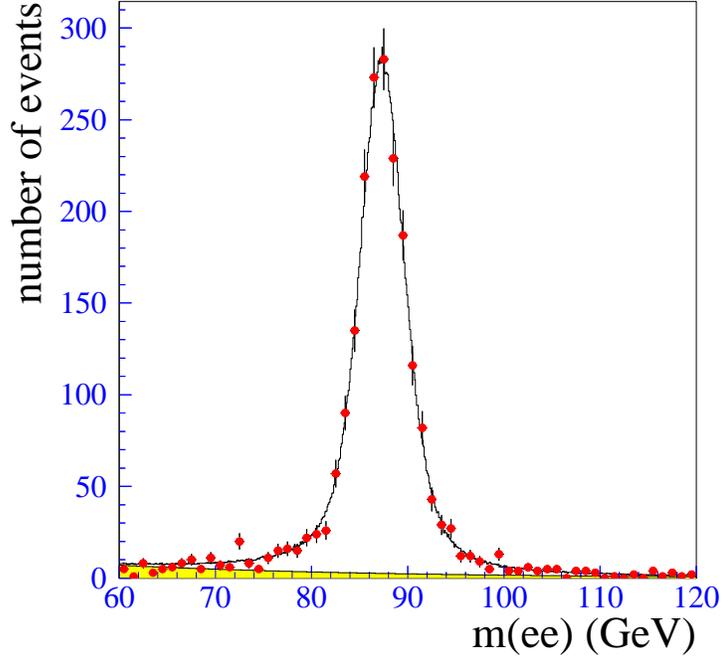,height=3.5in}
\caption{The resonant signal for a $Z$ boson via $Z\to \epem$
at the D0 detector.}
\label{Zee}
\end{figure}
\end{center}

As a simple example of a two-body decay, consider  $Z\to \epem$,
\be 
m_{ee}^2  = (p_{e^+} + p_{e^-})^2
\approx 2 p_{e^+} \cdot p_{e^-} \approx  2E_{e^+} E_{e^-} (1-\cos\theta_{\epem})
\approx M_Z^2,
\ee
which is invariant in any Lorentz frame, and leads to $E_e \approx M_Z/2$
in the $Z$-rest frame. Figure \ref{Zee} shows the peak 
in the $\epem$ invariant mass spectrum at $M_Z$, indicating the
resonant $Z$ production observed by the D0 collaboration \cite{d0} 
at the Tevatron collider.

Now let us examine the transverse momentum variable of a daughter
particle $p_{eT}^{}=p_e \sin\theta^*$, where $\theta^*$ is the polar angle
in the partonic c.m.~frame.  For a two-body final state kinematics, 
we thus have
\be
{d\hat\sigma\over dp_{eT}^{} } = {4p^{}_{eT} \over s \sqrt{1-4p_{eT}^2/s}}\ 
{d\hat\sigma\over d\cos\theta^* }.
\ee
The integrand is singular at  $p_{eT}^2 = s/4$, but it is integrable.
{
\vskip 0.2cm
\noindent
\tt Exercise: Verify this equation for  Drell-Yan production of $\epem$.
\vskip 0.2cm
}
\noindent
Combining with the Breit-Wigner resonance, we obtain
\be
{d\hat\sigma\over dm_{ee}^2\ dp_{eT}^2 } \propto
\frac{\Gamma_ZM_Z}{ (m^2_{ee}-M^2_Z)^2 + \Gamma_Z^2 M_Z^2}\ 
{1 \over m_{ee}^2\sqrt{1-4p_{eT}^2/m_{ee}^2}}\  {d\hat\sigma\over d\cos\theta^* }.
\ee
We see that the mass peak of the resonance leads to an enhanced
distribution near  $p_{eT}^{} = M_Z^{}/2$. This is called the Jacobian peak. 
This feature is present for any two-body kinematics with a fixed subprocess 
c.m.~energy. 

{
\vskip 0.2cm
\noindent
\tt Exercise: While the invariant mass distribution $d\sigma/dm_{\epem}$ is 
unaffected by the motion of the produced $Z$ boson, show that the 
$d\sigma/ dp_{eT}^{}$ distribution for a moving $Z$ with a momentum $\vec p_Z^{}$ is changed 
 with respect to a $Z$ at rest  at the leading order of $\vec \beta_Z^{} = \vec p_Z^{}/E_Z$.
\vskip 0.2cm
}

It is straightforward to generalize the invariant mass variable to multi-body
system. Consider a slightly more complicated signal of a Higgs decay
\be
H \to Z_1 Z_2 \to \epem\ \mu^+\mu^-.
\ee
Obviously, besides the two $Z$ resonant decays, the four charged
leptons should reconstruct the mass of the parent Higgs boson
\bea
m_H^2 &=& (\ \sum_i^4 p_{i} \ )^2 =  2 (M_Z^2 + p^{}_{Z_1} \cdot p^{}_{Z_2}) \\
 &=& (E_{e^+} + E_{e^-} + E_{\mu^+} + E_{\mu^-})^2
- (\vec p_{e^+} + \vec p_{e^-}  + \vec p_{\mu^+} + \vec p_{\mu^-})^2.
\label{Minv}
\eea

\subsubsection{The transverse mass variable}

\begin{center}
\begin{figure}[tb]
\psfig{figure=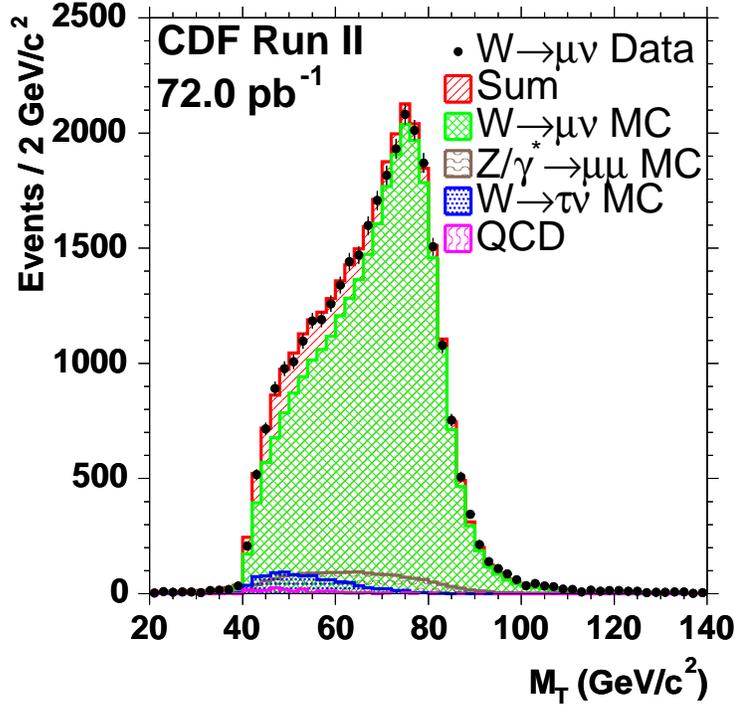,height=4in}
\caption{The transverse mass peak for a $W\to \mu\nu$ signal 
observed at CDF.}
\label{MTW}
\end{figure}
\end{center}

As another example of a two-body decay, consider  $W \to e\nu$.
The invariant mass of the leptonic system is
\be 
 m_{e\nu}^2 = (E_e + E_\nu)^2 - (\vec p_{eT} + \vec p_{\nu T})^2 
 - (p_{ez} + p_{\nu z})^2.
\ee
The neutrino cannot be directly observed by the detector and only
its transverse momentum can be inferred by the imbalancing 
of the observed momenta, 
\be
\vec{\ptmiss} = -\sum \vec {p_T}(observed), 
\ee
called missing transverse momentum, identified as 
$\ptmiss=p_{\nu T}$. Missing transverse energy is similarly defined,
and $\etmiss=E_\nu$. The invariant mass variable
thus cannot be generally reconstructed. We would get the correct value
of $m_{e\nu}$ if we could evaluate it in a frame in which
the missing neutrino has no longitudinal motion $p_{\nu z}=0$; but
this is impractical. Instead, one may consider to ignore the (unkown)
longitudinal motion of the leptonic system (or the $W$ boson) all
together, and define a transverse mass of the system \cite{MT}
\bea
 m_{e\nu T}^2 &=& (E_{eT} + E_{\nu T})^2 - (\vec p_{eT} + \vec p_{\nu T})^2 \\
&\approx& 2 \vec p_{eT} \cdot \vec p_{\nu T} \approx  2E_{eT} \etmiss\  (1-\cos\phi_{e\nu}),
\nonumber
\eea
where $\phi_{e\nu}$ is the opening angle between the electron and the neutrino
in the transverse plane. 
When a $W$ boson is produced with no transverse motion, 
 $E_{eT}= \etmiss = m_{e\nu T}/2.$
It is easy to see that the transverse mass variable is invariant under 
longitudinal boosts, and it reaches the maximum  $m_{e\nu T} = m_{e\nu}$, 
for $p_{ez} = p_{\nu z}$, so
that there is no longitudinal motion for the electron and the neutrino when
boosting to  the $W$-rest frame. In general, 
\be
0 \le m_{e\nu T} \le m_{e\nu} .
\ee
The Breit-Wigner resonance at $ m_{e\nu} = M_W$ naturally leads to
a kinematical peak near $m_{e\nu T} \approx M_W$ again due to the Jacobian
factor 
\be
{d\hat\sigma\over dm_{e\nu}^2\ dm^2_{e\nu, T} } \propto
\frac{\Gamma_WM_W}{ (m^2_{e\nu}-M^2_W)^2 + \Gamma_W^2 M_W^2}\ 
{ 1 \over m_{e\nu} \sqrt{ m_{e\nu}^2-m_{e\nu, T}^2}} .
\ee
In the narrow width approximation, $m_{e\nu T}$ is cut off sharply at  $M_W$.
In practice, the distribution extends beyond $M_W$ because of the finite
width $\Gamma_W$. This is shown in Fig.~\ref{MTW} 
as observed by the CDF collaboration \cite{cdf}
in the channel  $W\to \mu\nu$.

{
\vskip 0.2cm
\noindent
\tt Exercise: While the invariant mass distribution $d\sigma/dm_{e\nu}$ is unaffected 
 by the motion of the produced $W$ boson, show that the $d\sigma/ dm_{e\nu T}^{}$ 
 distribution for a moving $W$ with a momentum $\vec p_W^{}$ is changed with 
 respect to a $W$ at rest  at the next-to-leading order of $\vec \beta_W^{} = \vec p_W^{}/E_W$.
Compare this conclusion with that obtained for  $d\sigma/ dp_{eT}^{}$.
\vskip 0.2cm
}

\subsubsection{The cluster transverse mass variable}

A  transverse mass variable with more than two-body final state is
less trivial to generalize than the invariant mass variable 
as given in Eq.~(\ref{Minv}).  This is mainly due to the fact that for a system
with more than one missing neutrino, even their transverse 
momenta $p_{\nu_i T}$
cannot be  individually reconstructed in general,  rather,  only one value
$\ptmiss$ is experimentally determined, which is identified as the vector
sum of all missing neutrino momenta.  Thus the choice of the
transverse mass variable depends on our knowledge about
the rest of the kinematics, on how to cluster the other momenta, 
in particular realizing the intermediate resonant particles. 

\vskip 0.2cm
\noindent
\underline{$H\to W_1 W_2 \to q_1 \bar q_2\ e \nu$:}

As the first example of the transverse mass variable with multi-body
final state, let us consider a possible Higgs decay mode
to $WW$ with one $W$ subsequently decaying to $q_1 \bar q_2$ 
and the other to $e\nu$. Since there is only one missing neutrino,
one may construct the transverse mass variable in a straightforward 
manner according to the kinematics for the two on-shell $W$ bosons
\bea
\nonumber
 M_{T,WW}^2 &=&  (E_{T,W_1} + E_{T,W_2} )^2 - (\vec p_{T,W_1} + \vec p_{T,W_2} )^2 \\
&=& \left( \sqrt{p_{T,jj}^2 + M_W^2} + \sqrt{p_{T,e\nu}^2 + M_W^2}  \right)^2 
- (\vec p_{T,jje} + \vec{\ptmiss } )^2.
\nonumber
\eea
Note that if the decaying Higgs boson has no transverse motion (like being
produced via $gg$ fusion), then the last term vanishes $\vec{\ptmiss }=-\vec p_{T,jje}.$

However, this simple construction would not be suitable for a light Higgs boson
when one of the $W$ bosons (or both) is far-off shell. 
The above expression can thus be revised as 
\bea
 M_{T,WW}^{\prime 2} 
 = \left( \sqrt{p_{T,jj}^2 + m_{jj}^2} + \sqrt{p_{T,e\nu}^2 + m_{e\nu T}^2}  \right)^2 
- (\vec p_{T,jje} + \vec{\ptmiss } )^2.
\nonumber
\eea

Alternatively, one can consider to combine the observed two jets and a lepton 
together into a cluster, and treat the missing neutrino separately. One has 
\bea
 M_{T,WW}^{\prime\prime 2}
= \left(\sqrt{p_{T,jje}^2 + M_{jje}^2} + \ptmiss   \right)^2 
- (\vec p_{T,jje} + \vec{\ptmiss } )^2.
\nonumber
\eea
Which one of those $M_T$ variables is the most suitable choice for the signal
depends on which leads to the best Higgs mass reconstruction and a better
signal-to-background ratio after simulations.

\vskip 0.2cm
\noindent
\underline{$H\to Z_1 Z_2 \to \epem\ \nu\bar\nu$:}

In searching for this signal,  we define the cluster transverse mass based
on our knowledge about the $Z$ resonances  \cite{MTH},
\bea
\nonumber
 M_{T,ZZ}^2 &=&  (E_{T,Z_1} + E_{T,Z_2} )^2 - (\vec p_{T,Z_1} + \vec p_{T,Z_2} )^2 \\
&=& \left(\sqrt{p_{T,\epem}^2 + M_Z^2} + \sqrt{ {\ptmiss}^2 + M_Z^2}  \right)^2 
- (\vec p_{T,\epem} + \vec{\ptmiss } )^2.
\nonumber
\eea
If the parent particle ($H$) is produced with no transverse motion, then
$M_{T,ZZ} \approx 2 \sqrt{p_{T,\epem}^2 + M_Z^2}. $

{
\vskip 0.2cm
\noindent
\tt Exercise: Consider how to revise the above $M_{T,ZZ}$ construction
if $m_H^{} < 2M_Z$ in order to better reflect the Higgs resonance.
\vskip 0.2cm
}

\vskip 0.2cm
\noindent
\underline{$H\to W_1 W_2 \to \ell_1\nu_1\ \ell_2 \nu_2$:}

\begin{center}
\begin{figure}[tb]
\psfig{figure=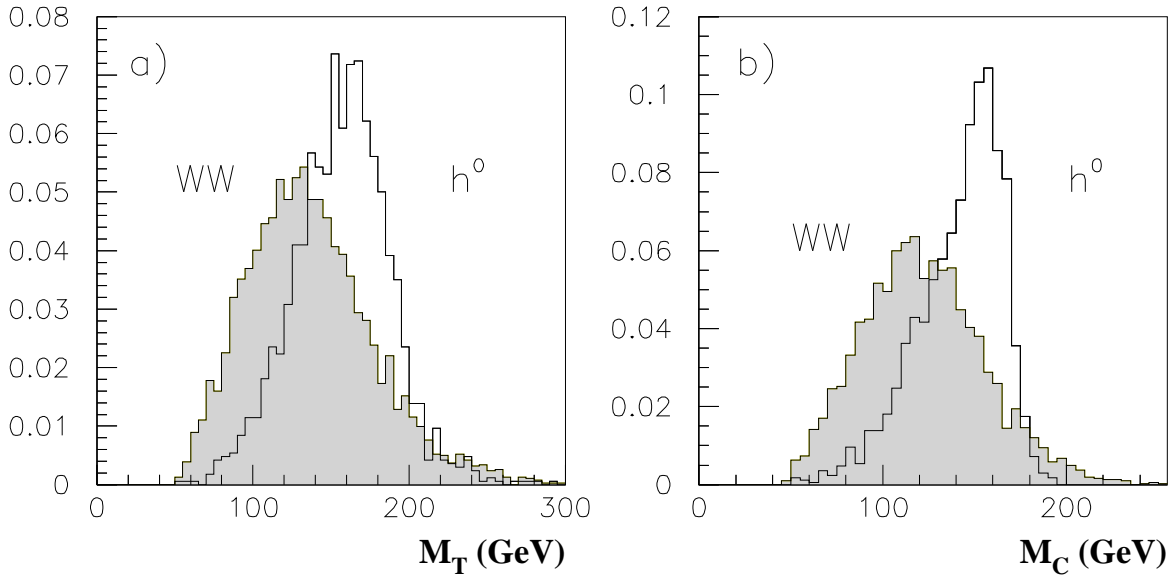,height=4.2in}
\vskip -1cm
\caption{ Normalized distributions 
${1\over \sigma} {d\sigma\over d M}$ for  $m_H^{}=170$ GeV (histogram)
and the leading $WW$ background (shaded) at the Tevatron for (a) $M_T$ 
in Eq.~(\ref{mt}) and (b) $M_C$ in Eq.~(\ref{mc}).}
\label{MTtc}
\end{figure}
\end{center}

As the last example, we consider a complicated case in which the two
neutrinos come from two different decays. The two missing neutrinos
do not present a clear structure, and thus one simple choice may be
to cluster the two charged leptons together  \cite{CMT}
\bea
 M_{C,WW}^{2}
= \left(\sqrt{p_{T,\ell\ell}^2 + M_{\ell\ell}^2} + \ptmiss   \right)^2 
- (\vec p_{T,\ell\ell} + \vec{\ptmiss } )^2.
\label{mc}
\eea
It was argued that since $\vec{\ptmiss} \approx - \vec p_{T,\ell\ell}$,
thus one should have, on average, $E_{T,\nu\nu} \approx E_{T, \ell\ell}$. 
This leads to a different construction \cite{Herbi}
\bea
 M_{T,WW}^{} \approx 2 \sqrt{p_{T,\ell\ell}^2 + M_{\ell\ell}^2}.
\label{mt}
\eea
These two choices are shown in Fig.~\ref{MTtc}, for $m_H^{}=170$ GeV
at the Tevatron, along with the SM $WW$ background \cite{HanZhang}.

\subsection{$t$-channel Enhancement: Vector Boson Fusion}
\label{EWA}

A qualitatively different process is initiated from gauge boson radiation,
typically off energetic fermions, as introduced in Sec.~\ref{WW}. 
Consider a fermion $f$ of energy $E$, the probability of finding a (nearly)
collinear gauge boson $V$ of energy $xE$ and transverse momentum
$\pt$ (with respect to $\vec p_f$) is approximated by \cite{sally,gordy}
\bea
\label{eq:T}
P^T_{V/f}(x,p_T^2)  &=& \frac{g_V^2+g_A^2}{ 8\pi^2}\
{ 1+(1-x)^2\over  x}\ { p_T^2\over (p_T^2+(1-x)M_V^2)^2}, \\
P^L_{V/f}(x,p_T^2) &=& \frac{g_V^2+g_A^2}{ 4\pi^2}\
{1-x\over  x}\ { (1-x)M_V^2\over (p_T^2+(1-x)M_V^2)^2},
\label{eq:L}
\eea
which $T\ (L)$ denotes the transverse (longitudinal) polarization of the
massive gauge boson. 
In the massless limit of the gauge boson, Eq.~(\ref{eq:T}) reporduces
the Weizs\"acker-Williams spectrum as in Eq.~(\ref{eq:WW}),
after integration of $p_T^2$ resulting in the logarithmic enhancement
over the fermion beam energy.  
In fact, the kernel of this distribution is the same as the quark 
splitting function $q\to qg^*$ \cite{esw,george}.

The scattering cross section can thus be formally expressed as
\bea
\sigma(f a \to f' X) \approx \int dx\ dp_T^2\ P_{V/f}(x, p_T^2)\  \sigma(V a\to X).
\label{ewa}
\eea
This is also called the Effective $W$ Approximation. Although this may
not be a good approximation until the parton energy reaches
$E\gg M_V$, it is quite important to
explore the general features of the $t$-channel behavior. 

First of all, 
due to the non-zero mass of the gauge boson, there is no more collinear
singularity. The typical value of the transverse momentum 
 (gauge boson or jet)  is $\pt \sim \sqrt{1-x}\ M_V \lsim M_W$. 
Since $x$ prefers to be low reflecting the infrared behavior, 
the jet energy $(1-x)E$ tends to be high.
These observations provide the arguments for ``forward jet-tagging" 
for massive gauge boson radiation processes: a highly energetic 
companion  jet with $\pt \lsim M_W/2$ and a scatteirng polar angle of a few 
degrees \cite{jettag}.  Furthermore, it  is very interesting to note 
the qualitative difference between  $P^T$ and $P^L$ for the $\pt$ dependence
(or equivalently the angular dependence) of the outgoing fermion.
For $\pt \ll M_W$, $P^T$ is further suppressed with respect to $P^L$;
while for $\pt > M_W$, $P^T$ is enhanced instead in the central scattering
region. This was the
original design for a forward jet-tagging and a 
``central jet-vetoing"  \cite{tagveto} to enhance the 
{\it longitudinal} gauge boson
fusion with respect to the {\it transverse} gauge boson fusion in the search
for strong $WW$ scattering signals \cite{baggeretal}. 
It has been further realized that the $t$-channel electroweak 
gauge boson  mediations
undergo color singlet exchanges, and thus do not involve significant QCD 
radiation. Consequently, there will be little hadronic activities connecting
the two parton currents. This further justifies the central jet-vetoing, and 
is developed into a ``mini-jet vetoing" to further
separate the gauge boson fusion processes from the large SM backgrounds
in particular those with QCD radiation in the central region \cite{dieter}.

\subsection{Forward-Backward Asymmetry}
The precision test  of universal chiral couplings of 
the gauge bosons  $Z^0,\ W^\pm$ to SM fermions is among the most
 crucial experimental confirmation for the validity of the SM.
Similar probe would be needed to comprehend any new vector bosons 
once they are discovered in future collider experiments.
The forward-backward asymmetry, actually the Parity-violation along the
beam direction, is very sensitive to the chiral structure of the vector boson 
to fermion couplings. 

Consider a parton-level process for a vector boson production and decay
\be
i\bar i \to V \to f \bar f,
\ee
where the initial state $i\bar i=e^-e^+,\ q\bar q$. 
Let us now parameterize the coupling vertex of a vector boson $V_\mu$ to 
an arbitrary  fermion pair $f$ by
\be
i \sum_\tau^{L,R} g_\tau^f \ \gamma^\mu\ P_\tau.
\ee
Then the parton-level forward-backward asymmetry is defined as
\be
A^{i,f}_{FB} \equiv  \frac{N_F - N_B}{N_F + N_B}
	= \frac{3}{4} \mathcal{A}_i \mathcal{A}_f,
\ee
where $N_F$ ($N_B$) is the number of events with the final-state fermion
momentum  $\vec p_f$ 
in the forward (backward) direction defined in the parton c.m.~frame
relative to the initial-state fermion  $\vec p_i$.  
The asymmetry $\mathcal{A}_f$ is given in 
terms of the chiral couplings as
\begin{equation}
	\mathcal{A}_f = \frac{(g_L^f)^2 - (g_R^f)^2}{(g_L^f)^2 + (g_R^f)^2}.
\end{equation}

The formulation so far is perfectly feasible in $\epem$ collisions. However,
it becomes more involved when applied to hadron colliders
$AB \to V X \to f \bar f X.$ The first problem
is  the mismatch between the parton c.m.~frame (where the scattering angle
is defined to calculate the asymmetry) 
and the lab frame (where the final-state fermion momentum is actually 
measured). This can be resolved if we are able to fully determine the final
state momenta $p_f,\ p_{\bar f}$.
We thus construct the vector boson momentum 
\be
p^{}_V = p_f + p_{\bar f}, 
\label{eq:pV}
\ee
and then boost $\vec p_f$ back to the $V$-rest frame, presumably the parton
c.m.~frame.
The second problem is the ambiguity of the $\vec p_i$ direction: both a quark 
and an anti-quark  as initial beams 
can come from either hadrons $A$ or $B$, making the 
determination of $\vec p_i$ impossible in general. Fortunately, one can
resolve this ambiguity to a good approximation. This has something to do
with our understanding for quark parton distributions in hadrons. We  first
recognize that for a heavy vector boson production, the parton energy
fraction is relatively large $x\sim M_V/\sqrt s$, and thus the contributions 
from valance quarks dominate, recall Fig.~\ref{fig:pdfs}. 
Consider the case at the Tevatron for $p \bar p$ collisions, 
we can thus safely choose the beam direction of the protons (with more 
quarks) as $\vec p_i$.
As for the LHC with $pp$ collisions, this discrimination of $p$ versus $\bar p$
is lost. However, we note that for $u\bar u,\ d\bar d$ annihilations, valance
quarks ($u,\ d$) carry much larger an $x$ fraction than the anti-quark in a proton.
We thus take the quark momentum direction $\vec p_i$ along with the
boosted direction as reconstructed in Eq.~(\ref{eq:pV}), 
recall the boost relation Eq.~(\ref{eq:rapcm}). With those clarifications, 
we can now define the hadronic level asymmetry at the LHC \cite{Rosner}
\begin{equation}
	A_{FB}^{\rm LHC} = 
	\frac{ \int dx_1 \sum_{q=u,d} A_{FB}^{q,f} 
	\left( P_q(x_1)P_{\bar q}(x_2) - P_{\bar q}(x_1)P_q(x_2) \right) 
	{\rm sign}(x_1-x_2)}
	{ \int dx_1 \sum_{q=u,d,s,c} 
	\left(P_q(x_1)P_{\bar q}(x_2) + P_{\bar q}(x_1)P_q(x_2) \right) },
	\label{eq:Afbcalc}
\end{equation}
where $P_q(x_1)$ is the parton distribution function for quark $q$ in the 
proton with momentum fraction $x_1$, evaluated at $Q^2 = M_V^2$.
The momentum fraction $x_2$ is related to $x_1$ by the condition 
$x_1x_2 = M_V^2/s$ in the narrow-width approximation.
Only $u$ and $d$ quarks contribute to the numerator since we explicitly 
take the quark and antiquark PDFs to be identical for the sea quarks;
all flavors contribute to the denominator. Some recent explorations on
the hadronic level asymmetry for various theoretical models 
have been presented in \cite{smoking}.


\subsection{Be Prepared for More Involved Inclusive Signatures}
\label{more}

%
%
%
%

The previous sections presented some basic techniques and general
considerations for seeking for new particles and interactions. 
They are applicable to many new physics searches. 
 Prominent examples include:
\begin{itemize}
\item Drell-Yan type of new particle production in 
$s$-channel \cite{zprime,LH2,LH1,smoking,TC,RS,herbi2}:
\bea
\nonumber
&& Z' \to \ell^+\ell^-,\ W^+W^-;\quad W' \to \ell \nu,\ W^\pm Z; \\
\nonumber
&& Z_H \to ZH;\quad W_H \to W^\pm H;\\
\nonumber
&& \rho^{0,\pm}_{TC} \to t\bar t,\ W^+W^-;\quad t\bar b,\ W^\pm Z; \\
\nonumber
&& {\rm heavy\ KK\ gravitons}\to \ell^+\ell^-,\ \gamma\gamma, ... ;\\
&& {\rm single}\ \tilde q,\  \tilde\ell\ {\rm via\ R\ parity\ violation}.
\nonumber
\eea
\item $t$-channel gauge boson fusion processes \cite{WWSUSY,dicus,LH1,Mike,smoking}:
\bea
\nonumber
&& W^+W^-,\ ZZ,\ W^\pm Z \to H,\  \rho^{0,\pm}_{TC},\ {\rm light\ SUSY\   partners};\\ 
\nonumber
&& W^+W^+ \to H^{++};\\ 
&& W^+ b \to T.
\nonumber
\eea
\end{itemize}

However, Nature may be trickier to us. Certain class of 
experimental signals for new physics at  hadron collider environments
may be way more complex than the simple examples illustrated above.
The following possible scenarios may make the new physics identification
difficult:
\begin{itemize}
\item A new heavy particle may undergo a complicated cascade decay, 
so that it is impossible to reconstruct its mass, charge etc. For example, think
about a typical gluino decay \cite{cascade} in SUSY theories 
$$\tilde g\to \bar q\ \tilde q \to \bar q\ q' \tilde\chi^+ \to \bar q\ q' \ \tilde\chi^0 W^+ 
\to \bar q\ q'  \ \tilde\chi^0\ e^+ \nu.$$
\item New particles involving electroweak interactions often yield weakly
coupled particles in the final state, resulting in missing transverse momentum
or energy, making it difficult for reconstructing the kinematics. 
Examples of resulting in missing energies 
include neutrinos in the SM, neutralino LSP in SUSY theories \cite{HaberKane}, 
light Kaluza-Klein gravitons in large extra dimension models \cite{GRW-HLZ},
and the lightest stable particles in other theories like in universal extra dimensions
(UED) \cite{UED} and little Higgs (LH) model with a T-parity \cite{T-P} etc.
\item Many new particles may be produced only in pair due to a conserved quantum
number, such as the R-parity in SUSY, KK-parity in UED, and T-parity in LH,
leading to a smaller production rate due to phase space suppression 
and more involved kinematics. 
For the same reason, their decays will most commonly yield a  final state
with missing energy. The signal production and the decay products are
 lack of characteristics.
\end{itemize}

On the other hand, one  may consider to take the advantage 
of those less common situations when identifying new physics signatures 
beyond the standard model. Possible considerations include:
\begin{itemize}
\item Substantial missing transverse energy is an important hint for new
physics beyond the SM, since $\etmiss$ in the SM  mainly comes from the
limited and predictable sources of $W,Z,t$ decays, 
along with potential poor measurements of jets.
\item High multiplicity of isolated high $\pt$ particles, such as multiple 
charged leptons and jets, may indicate the production and decay of
new heavy particles, rather than from higher order SM processes.
\item Heavy flavor enrichment is again another important feature
for new physics, since many  classes of new physics have enhanced
couplings with heavy flavor fermions, such as $H\to, b\bar b, \tau^+\tau^-$; 
$H^+ \to t\bar b, \tau^+\nu$; $\tilde H \to \tilde\chi H$;  
$\tilde t \to \tilde\chi^+ b, \tilde\chi^0 t$;
$\rho^{}_{TC}, \eta_{t}^{} \to t\bar t$ etc.
\end{itemize}
Clever kinematical variables may still be utilized, such as the lepton momentum
and invariant mass endpoints as a result of certain unique kinematics in SUSY
decays \cite{endpoint}.  
We are always encouraged to invent more effective  observables for new
signal searches and measurements of the model parameters. 

When searching for these difficult signals
in hadron collider environments, it is likely that we have
to mainly deal with event-counting above the SM expectation, without
``smoking gun" signatures. Thus it is of
foremost importance to fully understand the SM background 
processes, both for total 
production rates and for the shapes of kinematical distributions. 
This should be recognized as a serious challenge to theorists working
on collider phenomenology, in order to be in a good position for discovering 
new physics in hadron collider experiments.

To conclude these lectures, I would like to say  that it is highly anticipated
that the next generation of collider experiments at the LHC and ILC will
reveal exciting new physics beyond the currently successful standard
model. Young physicists should be well prepared for understanding the
rich but complex data from the experiments in connection to our 
theoretical expectation and imagination,  and thus contributing to the
major discovery. 
 
\begin{acknowledgments}
I would like to thank the TASI-2004 organizers, John Terning, Carlos Wagner,
and Dieter Zeppenfeld for inviting me to lecture,  K.T. Mahanthappa for his 
arrangements and hospitality,
and the participating students for stimulating discussions. I would also
like to express my gratitude to Yongsheng Gao, Yibin Pan, Heidi Schellman,
Wesley Smith, Weimin Yao, for their help in preparing the sections related to
the experimental issues. This work was supported in part
by the U.~S.~Department of Energy under contract No.~DE-FG02-95ER40896,
and by the Wisconsin Alumni Research Foundation.
\end{acknowledgments}

\appendix

\section{ Relativistic Kinematics and Phase Space }
\label{app-ps}

\subsection{Relativistic Kinematics}

Consider a particle of  rest mass $m$ and momenta $\vp$ 
moving in a frame $\cO$. 
We denote its four-momentum $p\equiv p^\mu=(E,\vp)$. 
The Lorentz invariant $m^2$ defines the on-mass-shell condition 
\bea
p^\mu p_\mu = E^2- \vec p \cdot \vec p = m^2.
\eea
Its {\it velocity} in units of $c$ is 
\bea
\vec\beta \equiv {\vec v\over c} = {\vec p\over E}\quad (-1 \le \beta \le 1);\quad
{\rm and}\quad \gamma \equiv (1-\beta^2)^{-{1\over2}} = {E\over m} .
\eea
Consider another frame $\cO'$ that is moving with respect to $\cO$ 
along the $\vec z$ direction (without losing generality). It 
is sufficient to specify the Lorentz transformation between the two frames
by either the relative  velocity  ($\beta_0$) of the moving $\cO'$
or its {\it rapidity}
\bea
\label{rapid}
y_0={1\over 2}\ln{1+\beta_0 \over 1-\beta_0 }\quad (-\infty < y_0 < \infty).
\eea
For instance, for the four-momentum vector
\bea
\nonumber
\label{trans}
\left(  
\begin{array}{l}
E' \\
 p'_z
\end{array}
\right)  & = &
\left(  
\begin{array}{ll}
\gamma_0 & -\gamma_0 \beta_0 \\
-\gamma_0 \beta_0 & \gamma_0
\end{array} 
\right) 
\left(  
\begin{array}{l}
E \\
 p_z
\end{array}
\right) \\
& = &
\left(  
\begin{array}{ll}
\cosh y_0 & -\sinh y_0 \\
-\sinh y_0 & \cosh y_0
\end{array}
\right)
\left(  
\begin{array}{l}
E \\
 p_z
\end{array}
\right),
\label{lor}
\eea
These transformations are particularly
useful when we need to boost the momentum of a decay product from the
parent rest frame ($\cO'$) to the parent moving frame ($\cO$). In this case,
the relative velocity is given by the velocity of the decaying particle
$\vec\beta = \vp^{parent}/E^{parent}$.

The Lorentz-invariant phase space element  for an $n$-particle final state 
can be written as 
\be
dPS_n \equiv (2\pi)^4\ 
\delta^4\left(P-\sum_{i=1}^n p_i \right) \Pi_{i=1}^n {1\over (2\pi)^3}
{d^3\vec p_i\over 2E_i} .
\ee
The $\delta^4$ imposes the constraint on the phase space by the four-momentum
conservation of the initial state total momentum $P$. 
Each final state particle satisfies an on-shell condition $p_i^2 = m_i^2$,
and the total c.m.~energy squared is $s=P^2 =\left(\sum_{i=1}^n p_i \right)^2.$

\subsection{One-particle Final State}
\label{one-body}
Most straightforwardly, we have the phase space element for one-particle final state
\be
dPS_1\equiv (2\pi)\ {d^3\vec p_1\over 2E_1} \delta^4 (P-p_1) \doteq
\pi | \vec p_1| d\Omega_1 \delta^3 (\vec P-\vec p_1),
\label{ps1}
\ee
here and henceforth, we adopt a notation $``\doteq"$ to indicate that certain less-concerned
variables have been integrated out at this stage. For instance, the variable $E_1$ has
been integrated out in the last step of Eq.~(\ref{ps1}), which leads to a trivial (but
important) relation $E^{cm}_1=\sqrt s$ in the c.m.~frame. 

Making use of the identity
\be
{d^3\vec p\over 2E} = \int d^4p\  \delta(p^2-m^2), 
\ee
we can rewrite the phase space element as
\be
dPS_1\doteq 2\pi\  \delta(s-m_1^2)= {\pi \over s}\  \delta(1-{m_1\over \sqrt s}).
\ee
We will call the coefficient of the phase-space element  ``phase-space volume", 
after integrating out all the variables. 
Here it is $2\pi$ for one-particle final state in our convention.

\subsection{Two-body Kinematics}
\label{two-body}

For a two-particle final state with the momenta $\vp_1,\ \vp_2$ respectively, 
the Lorentz-invariant phase space element  is given by
\bea
\nonumber
dPS_2 &\equiv&  {1\over (2\pi)^2}\ 
\delta^4\left(P- p_1-p_2 \right) {d^3\vec p_1\over 2E_1}  {d^3\vec p_2\over 2E_2} \\
 &\doteq & {1\over (4\pi)^2}\  {|\vp_1^{cm}|\over \sqrt s} \ d\Omega_1 
 = {1\over (4\pi)^2}\  {|\vp_1^{cm}|\over \sqrt s}\  d\cos\theta_1 d\phi_1.
\label{ps2}
\eea
Two-body phase space element is dimensionless, and thus no dimensionful 
variables unfixed.  That is to say that the two-body phase space weight is 
constant and the magnitudes of 
the energy-momentum of the two particles are fully
determined by the four-momentum conservation. 
It is important to note that the particle energy spectrum is monochromatic. 
Specifically, in the c.m.~frame
\bea
|\vp_1^{cm}| = |\vp_2^{cm}|={\lambda^{1/2}(s,m_1^2,m_2^2)\over 2\sqrt s},\ 
E_1^{cm}={s+m_1^2-m_2^2\over 2\sqrt s} ,\ 
E_2^{cm}={s+m_2^2-m_1^2\over 2\sqrt s} ,
\nonumber
\eea
where the ``two-body kinematic function" is defined as
\be
\lambda(x,y,z)=(x-y-z)^2-4yz=x^2+y^2+z^2-2xy-2xz-2yz,
\ee
which is symmetric under interchange of any two variables.
While the momentum magnitude is the same for the two
daughter particles in the parent-rest frame, the more massive
the particle  is, the  larger its energy is.

The only variables are the angles for the momentum orientation.
We  rescale the integration variables $d\cos\theta_1=2dx_1$
and $d\phi_1=2\pi dx_2$ to the range $0\le x_1, x_2 \le 1$, and thus
\bea
dPS_2  = {1\over 4\pi}{1\over 2}\ 
\lambda^{1/2}\left(1,{m_1^2\over s},{m_2^2\over s}\right) 
dx_1 dx_2.
\eea
It is convenient to do so in order 
to see the phase-space volume and to implement Monte Carlo simulations.

The phase-space volume of the two-body is scaled down 
with respect to that of the one-particle by a factor 
\be
{ \ dPS_2\over s\ dPS_1} \approx {1\over (4\pi)^2} .
\ee
Roughly speaking, the phase-space volume with each additional
final state particle (properly normalized by the dimensonful
unit $s$) scales down by this similar factor. It is interesting
to note that it is just like the scaling factor with each
additional loop integral.

It is quite useful to express the two-body kinematics by a set
of Lorentz-invariant variables.
Consider a $2\to 2$ scattering process $p_a + p_b\to p_1 + p_2$, 
the Mandelstam variables are defined as
\bea
\nonumber
s&=&(p_a+p_b)^2=(p_1+p_2)^2=E^2_{cm},\\
t&=&(p_a-p_1)^2=(p_b-p_2)^2=m_a^2+m_1^2-2(E_a E_1-p_ap_1\cos\theta_{a1}) ,\\
\nonumber
u&=&(p_a-p_2)^2=(p_b-p_1)^2=m_a^2+m_2^2-2(E_a E_2-p_ap_2\cos\theta_{a2}).
\eea
The two-body phase space can be thus written as
\be
dPS_2 = {1\over (4\pi)^2}\ 
{dt\ d\phi_1
\over s\ \lambda^{1/2}\left(1,{m_a^2/s},{m_b^2/ s}\right)}.
\ee

\noindent
{\tt Exercise: Assume that $m_a=m_1$ and  $m_b=m_2$. Show that 
\bea
\nonumber
t&=& -2 p_{cm}^2 (1- \cos\theta^*_{a1}) ,\\
\nonumber
u&=& -2 p_{cm}^2 (1+ \cos\theta^*_{a1}) + {(m_1^2-m_2^2)^2\over s},
\eea
where $p_{cm}=\lambda^{1/2}(s,m_1^2,m_2^2)/2\sqrt s$ is the momentum 
magnitude in the\\
 c.m.~frame.  This leads to $t\to 0$ in the collinear limit.}

\vskip 0.3cm
\noindent
{\tt Exercise: A particle of mass $M$ decays to two particles 
isotropically in its rest frame. What does the momentum distribution look
like in a frame in which the particle is moving with a speed $\beta_z$?
Compare the result with your expectation for the shape change for a
basket ball.
}

\begin{center}
\begin{figure}[tb]
\psfig{figure=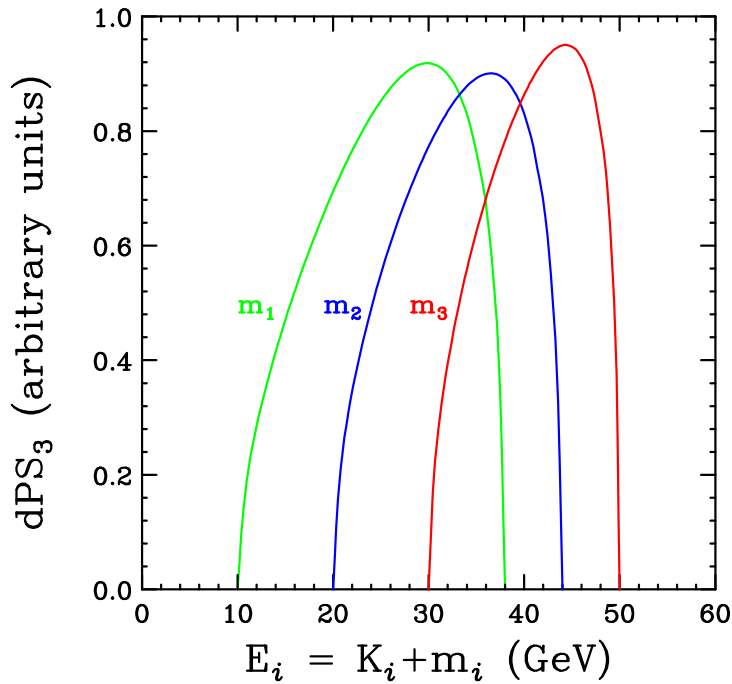,height=3.6in}
\caption{Three-body phase space weight as a function of $E_i\ (i=1,2,3)$
for $\sqrt s=100\ \gev,\ m_{1,2,3}=10,20,30\ \gev$, respectively.
\label{3body}}
\end{figure}
\end{center}

\subsection{Three-body Kinematics}
\label{three-body}
For a three-particle final state with the momenta 
$\vp_1,\ \vp_2,\ \vp_3$ respectively, 
the Lorentz-invariant phase space element  is given by
\bea
\nonumber
dPS_3 &\equiv&  {1\over (2\pi)^5}\ 
\delta^4\left(P- p_1-p_2-p_3 \right) {d^3\vec p_1\over 2E_1}
  {d^3\vec p_2\over 2E_2}   {d^3\vec p_3\over 2E_3} \\
 &\doteq & {|\vp_1|^2\ d|\vp_1|\  d\Omega_1 \over (2\pi)^3\ 2E_1}\ 
 {1\over (4\pi)^2}\  {|\vp_2^{(23)}|\over m_{23}} \ d\Omega_2\\
 &= & {1\over (4\pi)^3}\ 
  \lambda^{1/2}\left(1,{m_2^2\over m^2_{23}},{m_3^2\over m^2_{23}}\right)\ {2 |\vp_1| }\ 
 dE_1\  dx_2 dx_3 dx_4 dx_5.
\eea
The angular scaling variables are  $d\cos\theta_{1,2}=2dx_{2,4},$
and $d\phi_{1,2}=2\pi dx_{3,5}$ in  the range $0\le x_{2,3,4,5} \le 1$.
Unlike the two-body phase space, the particle energy spectrum is not
monochromatic. The maximum value (the end-point) for particle 1 in the 
c.m.~frame is 
\bea
E_1^{max}={s+m_1^2-(m_2+m_3)^2\over 2\sqrt s} ,\ \  {\rm or}\ \ 
|\vp_1^{max}| ={\lambda^{1/2}(s,m_1^2,(m_2+m_3)^2)\over 2\sqrt s}.
\eea
It is in fact more intuitive to work out the end-point for the kinetic energy
instead -- recall that this is how a direct neutrino mass bound is obtained
by examining $\beta$-decay processes \cite{PDG},
\bea
K_1^{max}= E_1^{max} - m_1 = 
{(\sqrt s-m_1-m_2-m_3) (\sqrt s-m_1+m_2+m_3) \over 2\sqrt s}.
\eea
Practically in Monte Carlo simulations, once $E_1^{cm}$ is generated
between $m_1$ to $E_1^{max}$, then all the other variables are determined
\bea
\nonumber
&& |\vp_1^{cm}|^2 = |\vp_2^{cm}+\vp_3^{cm}|^2 = (E_1^{cm})^2 - m_1^2, \\ 
&& m_{23}^2 = s - 2\sqrt s E_1^{cm} + m_1^2,\quad 
|\vp_2^{23}| = |\vp_3^{23}|={\lambda^{1/2}(m_{23}^2,m_2^2,m_3^2)\over 2 m_{23}},
\nonumber
\eea
along with the four randomly generated angular variables.

To see the non-monochromaticity of the energies in three-body 
kinematics, in Fig.~\ref{3body}, we plot the three-body phase space 
weight $dPS3$ as a function of $E_i\ (i=1,2,3)$.
For definiteness, we choose $\sqrt s=100$ GeV and $m_{1,2,3}=10,\ 20,\ 30$
GeV, respectively. It is in arbitrary units, but scaled to dimensionless by
dividing $E_i^{max} \sqrt s$. We see broad spectra for  energy distributions. 
Naturally, the more massive a particle is, the more energetic 
(energy and momentum) it is, but narrower for the energy spread.  
However, its kinetic energy $K_i=E_i-m_i$ is smaller for larger $m_i$.

\subsection{Recursion Relation for the Phase Space Element}
\label{rec}
\bea
\nonumber
dPS_n(P; p_1,...,p_n) &=& dPS_{n-1}(P; p_1,...,p_{n-1,n}) \\
                    && dPS_2(p_{n-1,n}; p_{n-1},p_n) {dm^2_{n-1,n}\over 2\pi}.
\eea

This recursion relation is particularly useful when we can write the intermediate 
mass integral for a resonant state. 

\begin{center}
\begin{figure}[tb]
\psfig{figure=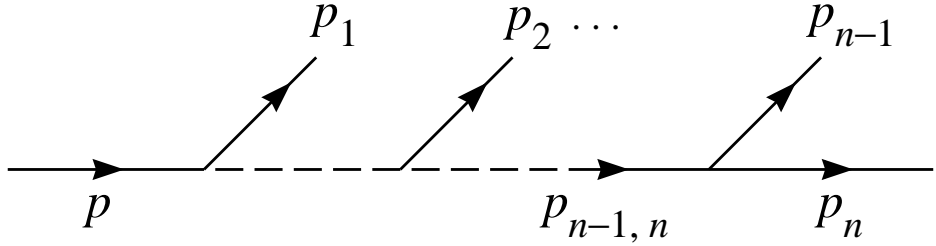,height=1.6in}
\caption{Illustration for the recursion relation for an $n$-body kinematics.
\label{fig:p}}
\end{figure}
\end{center}

\section{Breit-Wigner Resonance and the Narrow Width Approximation}
\label{sec:BW}

The propagator contribution of an unstable particle of mass $M$ and total width 
$\Gamma_V$ is written as 
\bea
R(s) = \frac{1}{ (s-M^2_V)^2 + \Gamma_V^2 M_V^2} ,
\label{BW}
\eea 
This is the Breit-Wigner Resonance. 

Consider a very general case of a virtual particle $V^*$ 
in an intermediate state, 
\be
a \to b V^* \to b\ p_1 p_2.
\ee
An integral over the virtual mass can be obtained by the reduction
formula in the last section. Together with kinematical considerations, 
the resonant integral reads
\be
\int_{(m^{min}_*)^2=(m_1+m_2)^2}^{(m^{max}_*)^2=(m_a-m_b)^2}
 {dm_*^2}.
\ee
The integral is rather singular near the resonance. Thus a variable
change is effective for the practical purpose,
\be
\tan\theta ={ {m_*^2 - M_V^2}\over {\Gamma_V M_V} },
\ee
resulting in a flat integrand over $\theta$
\be
\int_{(m^{min}_*)^2}^{(m^{max}_*)^2}
\frac{dm_*^2}{ (m^2_* - M^2_V)^2 + \Gamma_V^2 M_V^2} 
= \int_{\theta^{min}}^{\theta^{max}} \frac{d\theta }{\Gamma_V M_V} ,
\ee
where $\theta = \tan^{-1}(m_*^2 - M_V^2)/ \Gamma_V M_V$.
In the limit 
\be
(m_1+m_2)+\Gamma_V \ll M_V  \ll m_a - \Gamma_V,
\ee
then $\theta^{min} \to -\pi,\ \theta^{max}\to 0$.
This is the condition for the narrow-width approximation:
\be
\frac{ 1 }{ (m^2_* - M^2_V)^2 + \Gamma_V^2 M_V^2} \approx 
\frac{ \pi }{\Gamma_V M_V}\ \delta(m^2_* - M^2_V).
\ee
 
 {
\vskip 0.2cm
\noindent
\tt Exercise: Consider a three-body decay of a top quark, 
$t\to bW^* \to b\ e\nu.$
Making use of the phase space recursion relation and the narrow 
width approximation for the intermediate $W$ boson, show that the
partial decay width of the top quark can be expressed as
\be
\Gamma(t\to b W^* \to b\ e\nu) \approx \Gamma(t\to bW)\cdot BR(W\to e\nu).
\ee
\vskip 0.2cm
}


\end{document}